\documentclass[prapplied, amsmath,amssymbn,twocolumn,superscriptaddress]{revtex4-1}
\usepackage{graphicx}
\usepackage{color}
\usepackage[utf8]{inputenc}

\usepackage{blindtext}
\begin{document}

\title{Measuring transferability issues in machine-learning force fields: \\
The example of Gold-Iron interactions with linearized potentials}

\author{Magali Benoit}
\affiliation{CEMES, CNRS and Université de Toulouse, 29 rue Jeanne Marvig, 31055 Toulouse Cedex, France}
\author{Jonathan Amodeo}
\affiliation{Université de Lyon, INSA-Lyon, MATEIS, UMR 5510 CNRS, 69621 Villeurbanne, France}
\author{Ségolène Combettes}
\affiliation{CEMES, CNRS and Université de Toulouse, 29 rue Jeanne Marvig, 31055 Toulouse Cedex, France}
\author{Ibrahim Khaled}
\affiliation{Center for Nonlinear Phenomena and Complex Systems, Universit\'{e} Libre de Bruxelles, Code Postal 231, Boulevard du Triomphe, 1050 Brussels, Belgium}%
\author{Aurélien Roux}
\affiliation{Center for Nonlinear Phenomena and Complex Systems, Universit\'{e} Libre de Bruxelles, Code Postal 231, Boulevard du Triomphe, 1050 Brussels, Belgium}%
\author{Julien Lam}
\affiliation{Center for Nonlinear Phenomena and Complex Systems, Universit\'{e} Libre de Bruxelles, Code Postal 231, Boulevard du Triomphe, 1050 Brussels, Belgium}%
\email{julien.lam@ulb.ac.be}

\begin{abstract}
Machine-learning force fields have been increasingly employed in order to extend the possibility of current first-principles calculations. However,  the transferability of the obtained potential can not always be guaranteed in situations that are outside the original database. To study such limitation, we examined the very difficult case of the interactions in gold-iron nanoparticles. For the machine-learning potential, we employed a linearized formulation that is parameterized using a penalizing regression scheme which allows us to control the complexity of the obtained potential. We showed that while having a more complex potential allows for a better agreement with the training database, it can also lead to overfitting issues and a lower accuracy in untrained systems. 
\end{abstract}
\maketitle

\section{Introduction}

Atomistic modeling is often divided in two different types of simulations. On the one hand, quantum methods including Hartree-Fock and DFT approaches are considered the most accurate and are employed for virtually any types of chemical species\citep{marx2009ab,martin2008electronic}. On the other hand, classical force fields are used to perform large-scale and long-time simulations with less accuracy\cite{Frenkel2002,stone2016the}. However, it is still difficult to connect both approaches and until now, one can hardly perform a simulation involving millions of atoms for nanoseconds while retaining the accuracy of quantum methods. 

In this context, machine-learning interaction potentials (MLIP) have been proposed in the recent years and have shown great potentials to achieve such simulations\cite{Behler2015Aug,Bartok2015Aug,Behler2016Nov}. Numerous approaches are currently considered including Artificial Neural Networks\cite{Behler2007Apr}, Gaussian approximation methods\cite{Bartok2010Apr}, Linearized potentials\cite{Seko2014Jul,Goryaeva2019Aug}, Spectral Neighbor Analysis Potential\cite{Thompson2015Mar}, Symmetric Gradient Domain Machine learning\cite{Chmiela2017May,Chmiela2018Sep} and Moment Tensor Potentials\cite{Shapeev2016Sep}. The success of these techniques is recognized by the large variety of materials that were successfully tackled: pure metals\cite{Novoselov2019Jun,Seko2015Aug,Takahashi2018Jun,Zeni2018Jun,Botu2017Jan}, organic molecules\cite{Bereau2018Jun,Sauceda2019Mar,Bartok2017Dec,Veit2019Apr}, oxides\cite{Artrith2011Apr,Quaranta2017Apr}, water\cite{Nguyen2018Jun,Bartok2013Aug,Morawietz2012Feb,Natarajan2016Oct,Morawietz2013Aug}, amorphous materials\cite{Deringer2017Mar,Bartok2018Dec,Caro2018Apr,Deringer2018Jun,Deringer2018Nov,Sosso2018Jul} and hybrid perovskites\cite{Jinnouchi2019Jun}.

For all of these techniques, the main procedure consists in using a very universal analytical formulation for the force field which is then parameterized to match a database of DFT calculations including total energy, forces and stress tensors. However, it is admitted that MLIP can sometimes show poor transferability towards systems that are not included in the learning database. In the worst scenario, the MLIP is so-well fitted to its learning database that non-physical behaviors may be observed outside of it. In order to fix this issue, the main proposal is to regularly check the accuracy of the potential as the machine-learning molecular dynamics simulations are carried out and to improve  the MLIP "on the fly"\cite{Jinnouchi2019Jun,Jinnouchi2019Jul,Li2015Mar}. Yet, to the best of our knowledge, such flaw of the approach has never been quantitatively investigated while being acknowledged by both users and developers. 

For our case study, we choose interactions in gold-iron nanoparticles. In principle, such system can be found concurently in three different chemical orderings namely alloy, Janus and core-shell. Yet, recent experiments have shown that the synthesized Au-Fe nanoparticles are made of an iron core wrapped in a gold shell and that the shape of the iron core depends strongly on the amount of surrounding gold\cite{Langlois2015Aug,Benzo2019Sep,Tymoczko2019,Tymoczko2018Sep,Ponchet2020Aug}. These nanoparticles have potential biomedical applications as iron is known for its intrinsic ferromagnetism and gold capping can protect the iron core from oxidation. However, rationalizing the results of the synthesis along with predicting the material properties would require numerical simulations which are sparse for gold-iron  nanoparticles\cite{Vernieres2019Jul,Calvo2017Mar,Hong2015Oct,Combettes2020Sep,Ponchet2020Aug}. Indeed, while full quantum calculations can not be employed to study clusters of more than tens of atoms, the empirical modeling of gold-iron nanoparticles is a also very difficult case because these two metals are non miscible at room temperature on a very large domain of the phase diagram. There are therefore no iron-gold alloy crystal structures on which to adjust the parameters of an empirical potential. Moreover, iron is magnetic and crystallizes in a bcc structure while gold crystallizes in an fcc structure, which makes the development of a potential capable of capturing all the properties of this alloy even more complex. Previous attempts have shown their limits by stabilizing metastable alloys\cite{Zhou2004Apr} or by not being able to find the most stable Fe/Au interfaces\cite{Calvo2017Mar,Vernieres2019Jul}, leading us to develop potentials specifically dedicated to a particular problem, and hence highly non-transferable\cite{Combettes2020Sep}.

In this article, we begin by describing the methodology including linearized machine-learning potential and a penalizing regression scheme. In the results section, we first studied the influence of the descriptor functions. Then, we showed that the methodology allows one to quickly obtain MLIPs with different degrees of complexity. Afterwards, the transferability of these different potentials was tested on forces in untrained chemical orderings namely Janus and Core-shell. While the error should decrease monotonically when increasing the MLIP complexity, we observed a surprising non-monotonic behavior thus illustrating that more complexity does not necessarily lead to a better MLIP overall. Such transferability issue was reduced by using a more diverse set of descriptors. Finally, we measured some properties of the bulk and investigate the possibilities and the limitations of the obtained MLIP towards bulk simulations even if it was trained on nanostructures. 

\section{Methods}

\subsection{The $\Phi$-Lassolars machine-learning interaction potential}

For our MLIP, we employed the analytical formulation originally put forward by Seko et al.\cite{Seko2015Aug,Seko2014Jul,Takahashi2018Jun,Takahashi2017Nov,Seko2019Jun}. In this method, the total potential energy of a configuration made of $N$ atomic positions is first given by $E_{tot}=\sum_{i=0}^N E_{i}$ where $E_i$ is the atomic energy. For $E_i$, we considered a weighted linear combination of descriptors indexed by $n$: 
\begin{equation}
E^{(i)}=\sum_n \omega_n X_n^{(i)}
\label{Etot}
\end{equation}
where $\omega_n$ is the linear coefficient associated with the descriptor $X_n^{(i)}$. Until now, moment tensors\cite{Novoselov2019Jun}, group-theoretical high-order rotational invariants\cite{Seko2019Jun} and bispectrum components\cite{Wood2018Jun,Thompson2015Mar,Goryaeva2019Aug} were previously proposed as descriptors for such linearized potentials. In this work, we favored a simpler formulation which consists in developing the descriptor space in explicit two-body, three-body and N-body interactions:
\begin{eqnarray}
\left[\textrm{2B}\right]_{n}^{(i)}&=&\sum_j f_n(R_{ij}) \times f_c(R_{ij})\\ 
\left[\textrm{3B}\right]_{(n,l)}^{(i)}&=&\sum_j \sum_k  f_n(R_{ij})f_c(R_{ij})  f_n(R_{ik})f_c(R_{ik}) \cos^l(\theta_{ijk})\\ 
\left[\textrm{NB}\right]_{(n,m)}^{(i)}&=& \left[ \sum_j f_n(R_{ij}) \times f_c(R_{ij})  \times f_s(R_{ij}) \right] ^m
\end{eqnarray}
where $R_{ij}$ is the distance between atoms $i$ and $j$, $\theta_{ijk}$ is the angle centered around the atom $i$, and $l$ and $m$ are two positive integers. For the cut-off function, we chose what was originally proposed by Behler and Parrinello\cite{Behler2007Apr}: $f_c=\frac{1}{2}\left(1+\cos(\pi(R_{ij}/R_{cut})) \right)$ with $R_{cut}$ being set at $6$\,\AA. The switch function denoted $f_s$ is employed in order to prevent from non-physical behavior of the N-body contribution at short distances. To do so, two distances $r_1$ and $r_2$ are first defined as respectively 95\% and 105\% of the minimum of the dimer interactions and then a function is constructed to smoothly go to from $0$ to $1$ in the range of $[r_1:r_2]$: $f_s(u)=6u^5-15u^4+11^3$ where $u=(R_{ij}-r_1)/(r_2-r_1)$\cite{Wang2019Jun}. Altogether, these expressions allow for a direct computation of the forces as well as the stress tensors by differentiating with respect to the positions.\cite{Seko2014Jul}  Regarding the basis of functions $f_n$, there are no physical restrictions. In particular, for the two-body interactions, one can tune $f_n$ to mimic traditional interatomic potentials as for example Morse, Lennard-Jones, Buckingham or Yukawa potentials or use simple functions like Gaussians, Lorentzian or Asymmetric log-normal functions. Likewise, for the three-body interactions, the current formulation is very similar to what is done in the Stillinger-Webber potential\cite{Stillinger1985Apr}. Finally, the N-body interaction is a generalized form of the EAM potential where the embedding function is a polynomial of the atomic density\cite{Daw1984Jun}. In this case, the integer $m$ corresponds directly to the degree of N-body order. The difference between our formulation and the most recent ones proposed by Seko et al. is that, in the N-body interactions, we did not include any explicit angular dependence and did not mix different forms of the atomic density. 

For the fitting procedure of such a linear model, previous studies have proposed the use of genetic algorithm\cite{Thompson2015Mar,Wood2018Jun}, weighted ordinary least squares\cite{Goryaeva2019Aug}, Bayesian linear regression\cite{Jinnouchi2019Jul}, ridge regression\cite{Takahashi2017Nov,Takahashi2018Jun,Seko2019Jun} and Least absolute shrinkage and selection operator (Lasso)\cite{Seko2015Aug,Seko2014Jul}. In order to construct a simpler MLIP, we employed the Lasso regression with the Least Angle Regression Scheme (Lars, together denoted LassoLars)\cite{Efron2004Apr}. In practice, along with the ordinary least square objective function, $\chi^2_{OLS}$, the Lasso scheme adds a penalty on the sum over the absolute value of the coefficients $\omega_n$ and the employed error function is therefore given by:
\begin{equation}
\chi^2=\chi^2_{OLS}+\alpha\sum_{n} \vert\omega_{(n)} \vert
\end{equation}
where $\alpha$ is a parameter that controls the degree of penalty. The penalty on the absolute value of the coefficients enforces lots of the linear coefficients  to be exactly $0$. Additional, using Lars allows us to select the most relevant descriptors by measuring their correlation to the target. Using LassoLars as a regression scheme is at the expense of accuracy and flexibility for the MLIP but it allows for a considerable reduction in the complexity of the potential [Please see the Appendix for further motivation regarding the choice of LassoLars and for additional details on the $\Phi$-LassoLars implementation].

\subsection{DFT database}

\begin{figure}[ht!]
\includegraphics[width=8.6cm]{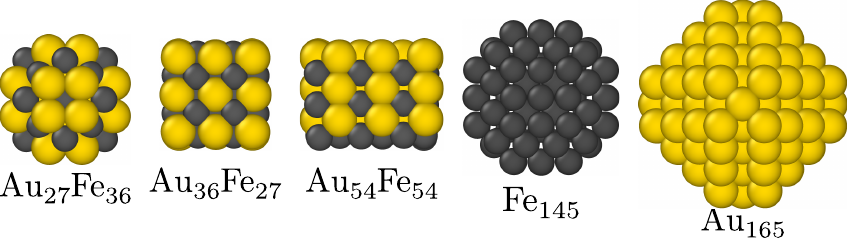}
\caption{Images of the initial structures employed in the database.}
\label{DFT}
\end{figure}

Building a general purpose potential for Au-Fe nano-crystals would require to model the atomic interactions not only in different structures (crystal polymorph, interfaces and liquid for instance) but also in different chemical orderings including alloyed, Janus and core-shell nanoparticles. Because the focus of this work is to measure the issues related to transferability, we purposely employed an incomplete database made of only three types of nanoparticles: (1) Alloys with almost equimolar compositions, (2) Pure iron in the body-centered cubic (bcc) phase, (3) Pure gold in the face centered cubic (fcc) phase[see Fig.\,\ref{DFT}]. Then, molecular dynamics (MD) simulations by means of a house code were carried out to melt the constructed nanoparticles. Simulations were performed in the NVT ensemble obtained with Andersen thermostat at 1400\,K during 500 ps using a timestep of 1 fs. We used simple pair-wise potentials made of Lennard-Jones and Morse interactions for gold and iron respectively and of Lennard-Jones interaction for the gold-iron cross-interaction. The employed Morse parameterization is the one of Hung et al.\cite{MaP} while the two Lennard-Jones potentials for gold-gold and iron-gold were simply parameterized in order to match the bulk lattice parameters and the cohesive energy. Along the melting path, we extracted configurations that are representative of the solid to liquid transition. In addition, each initial structure was also manually compressed. The distances are reduced by a factor of 75\% and along the compression, we extracted ten configurations in order to sample structures of higher density and better reproduce the repulsion at short distances. For the same reason, diatomic molecules FeFe, AuFe and AuAu with distances down to $1$\AA were also added in the database. For each of these configurations, forces were finally computed at the DFT-level using single-point calculations. Spin-polarized DFT calculations were performed with the VASP code \cite{kresse_efficiency_1996}, using PAW type pseudopotentials for iron and gold\cite{blochl_projector_1994}, a plane wave cutoff of 650 eV and a Methfessel-Paxton smearing parameter $\sigma$ of 0.01 eV. All calculations were done at the $\Gamma$-point of the Brillouin zone. Altogether, the database is made of $181653$ atomic configurations with an almost equal proportion in the three types of nanoparticles (34\% of alloy, 34\% of pure gold and 32\% of pure iron). We note that the database sampling was made with classical interaction potentials which is less satisfying than performing ab initio MD. Yet, the advantage is that it allows us to quickly sample configurations that remain physically valid and it prevents from running very computationally demanding simulations for those relatively large nanoparticles (more than $50$\,atoms and more than $100$ for respectively the alloy and the pure nanoparticles). In addition, by employing configurations that span from the crystalline to the liquid regime, we assure a large variety of atomic neighborhoods with forces ranging from $10^{-4}$ to $5$\,eV/\AA.  Finally, during the manual compression of the nanoparticle and when computing the forces for diatomic molcules, it may happen that some atoms are very close thus leading to very large forces. In the $\chi^2$, those large forces would contribute a lot while being very unlikely to emerge in a realistic dynamics. Therefore, before performing the fitting procedure, the database is filtered out to remove cases where the forces are lower than $5$\,eV/\AA. Such value was chosen large enough to keep some part of the repulsive interaction but not too large to avoid a fitting that is focused only on the repulsive part. Within the fitting procedure, the database is randomly divided in a training (95\%) and a test (5\%) sets. 

\section{Results}

\subsection{Influence of the descriptor space}

\begin{figure}[ht!]
\includegraphics[width=8.6cm]{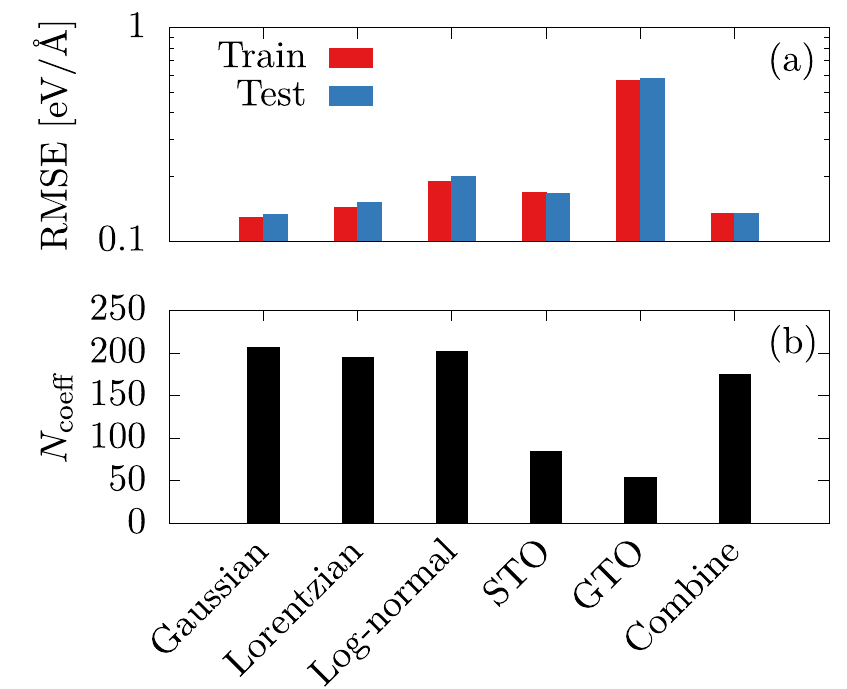}
\caption{(a) RMSE obtained for five different descriptors and measured on the training and the test sets. (b) Number of non-zero coefficients obtained with $\alpha=10^{-7}$}
\label{VsFunc}
\end{figure}

First, five different types of descriptors were tested. In particular, we used three functions that are "peak" functions ie. Gaussian, Lorentzian and Log-normal peaks and two functions that are usually employed for orbital calculations and that diverge at short distance ie.  Slatter-type (STO) and Gaussian-type (GTO) orbital [Please see Table\,I]. Then, the LassoLars method is employed to obtain five different interaction potentials using $\alpha=10^{-7}$. For the three-body and the N-body interactions, we used respectively $l=[1, 2, 3, 4, 5]$ and $m=[4, 5, 6, 7]$. Altogether, we have around a thousand available descriptors which include Au-Au, Fe-Fe and Fe-Au interactions for each type of descriptors. In Fig.\,\ref{VsFunc}.a, we show the fitting error measured as the root mean square error (RMSE) for the five different types of descriptors. RMSE measured on training and test data sets are similar which means that at this stage, no overfitting is observed. In addition, it appears that the two orbital functions are not as good as the peak functions although the STO still give an RMSE equal to $0.17$\,eV/\AA. Gaussian and Lorentzian  descriptors are able to reproduce the forces with an RMSE respectively equal to $0.13$ and $0.14$ \,eV/\AA. Such values are similar to what is obtained with the generally employed MLIP methods including neural networks, Gaussian approximation method and linearized potentials\cite{Zuo2020Jan}. In Fig.\,\ref{VsFunc}.b, the number of non-zero coefficient is plotted for the five different types of descriptors. In the cases of GTO and STO functions, much fewer descriptors were selected in comparison to the peak functions. In overall, it remains that the LassoLars algorithm allows one to drastically decrease the number of employed descriptors with respect to the number of available descriptors.

\begin{table*}[]
\addtolength{\tabcolsep}{0.5cm}
\begin{tabular}{|l|c|c|r|r|}
\hline Function name & Equation & List of ${a_n}$ & List of ${b_n}$ &  $N_{\mathrm{func}}$ \\ \hline\hline 
Gaussian &  $f_n(R_{ij})=\exp(-a_n(R_{ij}-b_n)^2)$ &  $0.5, 1.0, 1.5$ & $1, 2, 3, 4, 5, 6$ & $990$ \\
Lorentzian peak & $f_n(R_{ij})=1/((R_{ij}-a_n)^{b_n}+1)$.  & $1, 2, 3, 4, 5, 6$  & $2, 4, 6, 8$ & $1320$ \\
Log-normal peak & $f_n(R_{ij})=\exp(-\ln [(R_{ij}-a_n)/b_n]^2 )$ & $1, 2, 3, 4, 5, 6$ &   $1.0, 1.5, 2.0, 2.5$ & $1320$ \\
Slatter-type orbital &  $f_n(R_{ij})=R_{ij}^{a_n} \exp(-b_nR_{ij})$ &  $-2, -1, 0, 1, 2$ & $2, 4, 6, 8, 10$ &  $1375$ \\
Gaussian-type orbital & $f_n(R_{ij})=R_{ij}^{a_n} \exp(-b_nR_{ij}^2)$ &  $-2, -1, 0, 1, 2$ & $2, 4, 6, 8, 10$  &   $1375$ 
\\\hline
\end{tabular}
\label{TableFN}
\caption{Summary of the tested descriptors.}
\end{table*}

\begin{figure}[t!]
\includegraphics[width=8.6cm]{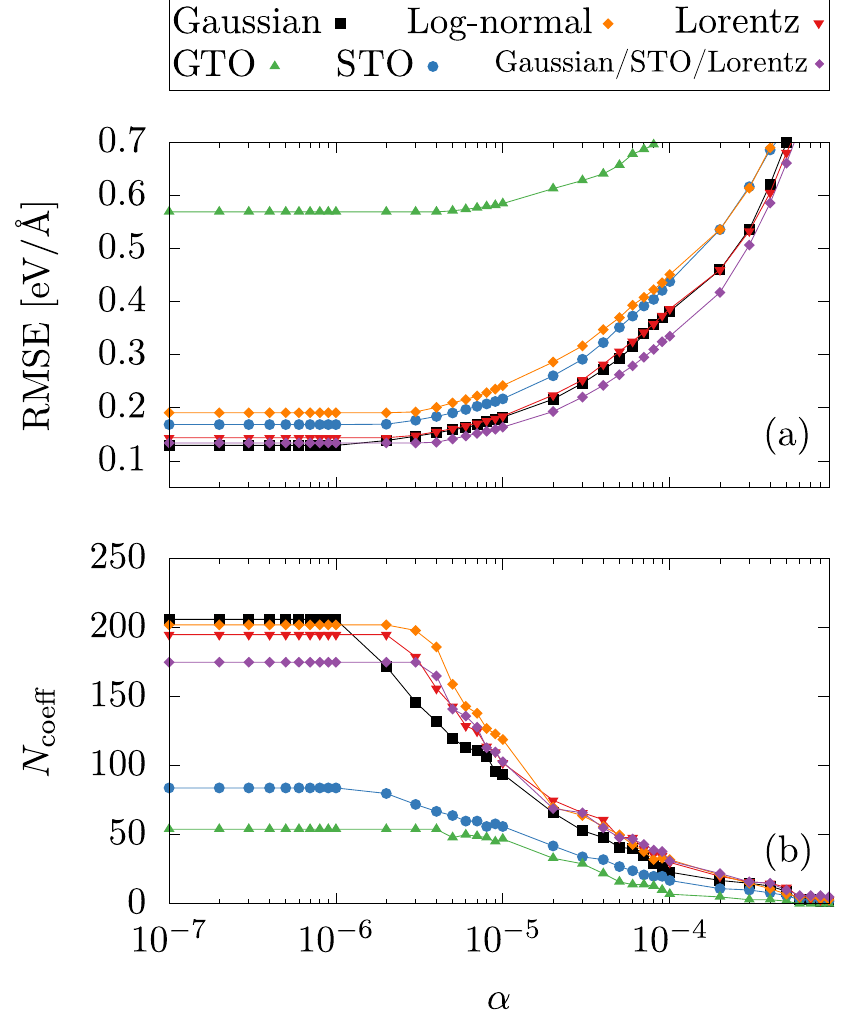}
\caption{Influence of the penalizing factor $\alpha$ on testing RMSE and on (b) the number of non-zero coefficients.}
\label{Alpha}
\end{figure}

In addition, before studying the transferability issues, we wish to illustrate a second advantage of using the LassoLars algorithm which consists in having a penalizing parameter that controls both the accuracy and the complexity of the obtained potential. According to Fig.\,\ref{Alpha}, by increasing $\alpha$, the number of non-zero coefficients and the computational cost can be reduced at the expense of increasing the RMSE. As such, with the LassoLars algorithm, one can simply choose which degree of accuracy or complexity is required for their usage. Finally, the presence of a plateau for the smallest values of $\alpha$ shows that the LassoLars regression only selects relevant descriptors thus reducing the potential complexity. Similar to what was obtained previously using $\alpha=10^{-7}$, we note that: (1)\,the 3 peak functions are the most accurate and behave similarly and (2)\,the STO function gives slightly higher RMSE yet with much fewer non-zero coefficients. Fig.\,\ref{Alpha} evidences that the LassoLars algorithm gives the ability to finely control the complexity of the potential at the expense of the accuracy on the DFT database. In the following, we will test how this complexity can influence the MLIP transferability.

\subsection{Complexity vs Transferability}

\begin{figure}[t!]
\includegraphics[width=8.6cm]{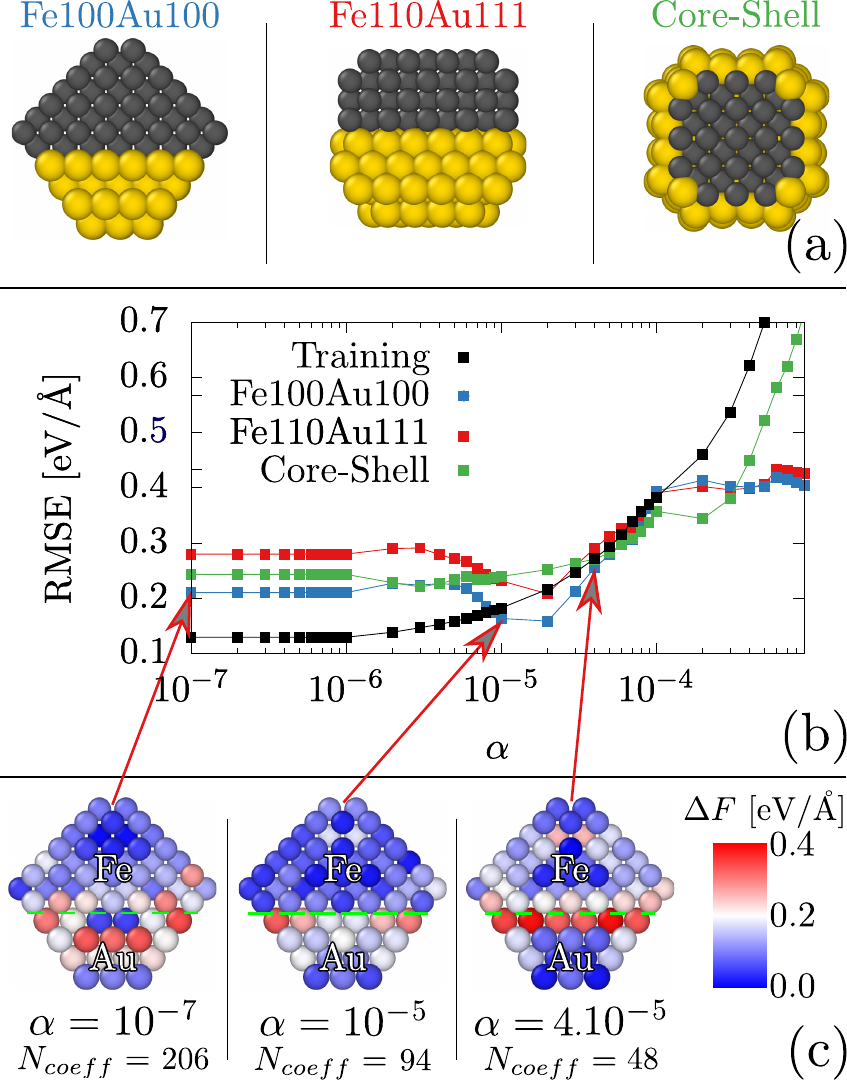}
\caption{(a) Images of untrained configurations with morphology Janus and Core-shell,
(b) Fitting error as a function of fitting parameter $\alpha$ for each untrained configurations and for the training set and (c) Map of the force errors on the Fe100Au100 Janus structure for different values of $\alpha$. The green line designates the iron/gold interface.}
\label{Trans}
\end{figure}

For that purpose, three additional morphologies of gold-iron nanoparticles were designed: two Janus and one core-shell [see Fig.\,\ref{Trans}.a]. Being able to accurately retrieve the interactions in those structures is a difficult test for the MLIP as the training set did not posses any of those demixed structures. Fig.\,\ref{Trans}(b) shows the corresponding RMSE on the forces without having trained the potential on these structures and using only Gaussian functions. Fig.\,\ref{Trans}(c) shows that most of the errors are located at the gold/iron interface which was not included in the training database. Surprisingly, the RMSE behavior is non-monotonic with a minimum located for the three structures around  $\alpha=8\times 10^{-6}$ which corresponds to $90$ non-zero coefficients. More specifically, when transferring the obtained potential to Fe100Au100 Janus nanoparticles, the RMSE obtained with $\alpha=10^{-7}$ is 30\% higher than with the less complex potential that was obtained  with $\alpha=8\times 10^{-6}$. Therefore, while increasing the complexity of the potential leads to a better agreement within the training database, it does not necessarily lead to an improvement of the RMSE in unlearned  structures using in our case different chemical ordering (ie. Janus and Core-shell instead of alloy nanoparticles). Our challenging test demonstrates that precautions should be made when using machine-learning approaches and that increasing the complexity does not automatically lead to a better overall potential. 

\begin{figure}[t!]
\includegraphics[width=8.6cm]{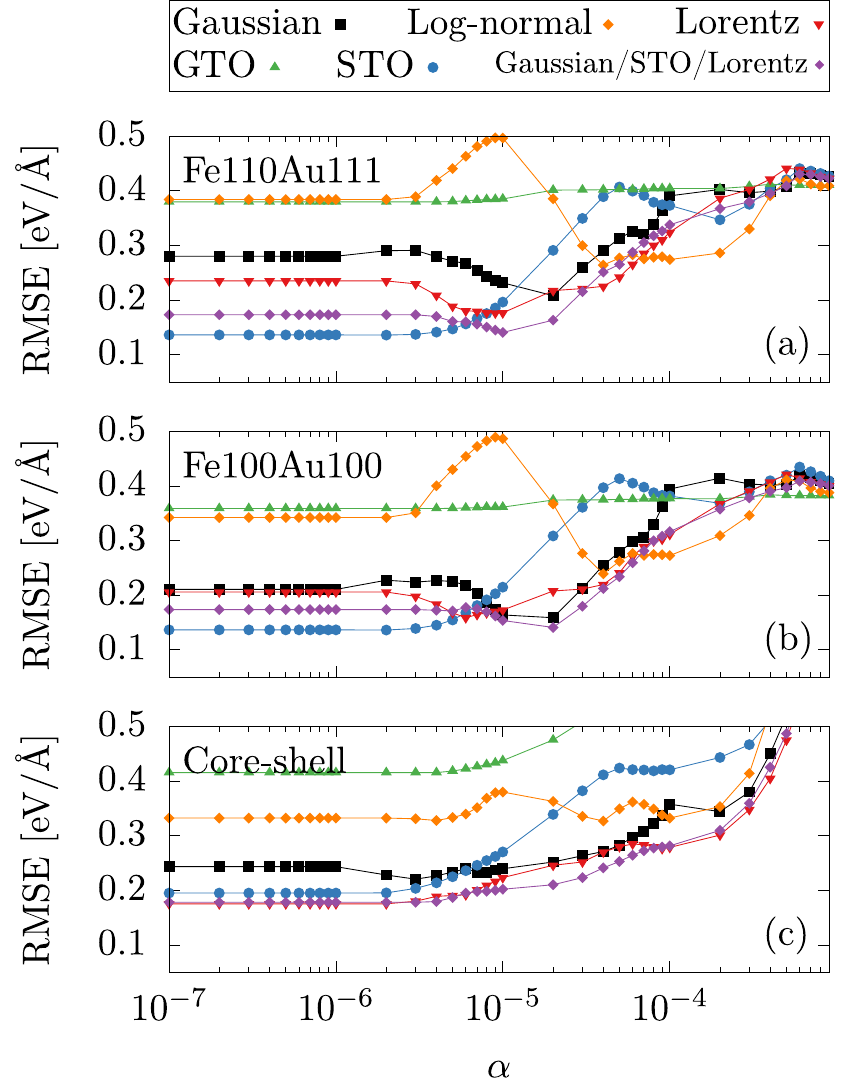}
\caption{(a) Number of non-zero coefficients and (b) RMSE as a function of the constrain parameter $\alpha$ when using simultaneously Gaussian, STO and GTO functions as descriptors. The dotted lines correspond to results when using only Gaussian functions \textit{ie.} results from Fig.\,\ref{Trans}b.}
\label{MoreFN_1}
\end{figure}

According to Fig.\,\ref{MoreFN_1}, the non-monotonic behavior that was highlighted when using only Gaussian functions is also observed with the two other peak functions ie. Lorentzian and Log-normal. We also note that, GTO is again very inaccurate and should most probably be avoided for usage as descriptor. Finally, the STO functions while being less accurate on the training database seems to give in overall better results in terms of transferability.

Finally, a combination of three different descriptors (Gaussian, Gaussian-type orbital and Lorentzian peak) is tested in order to employ simultaneously two different types of peak functions along with a function that diverge at short repulsion. STO was chosen for its remarkable ability to decrease the number of non-zero coefficient while Lorentzian peaks also showed slightly less non-monotonocity. On the one hand, regarding its fitting performance, this combination gives an RMSE similar to that obtained previously yet with fewer non-zero coefficients [See Fig.\,\ref{MoreFN_1}]. Having these two additional functions gives more flexibility in the descriptor space and fewer functions can therefore be selected for the same accuracy. On the other hand, for the transferability towards the unlearned structures, the combination gives RMSE that are comparable to the peak functions and even better in two cases (ie. Fe110Au111 and Core-Shell). More importantly, it allows for a further reduction of the non-monotonicity. We note that being able to combine three different types of descriptors at the same time is an additional advantage of using a constrained linear regression scheme such as LassoLars.

\subsection{Quality of the potential}

\begin{figure}[h!]
\includegraphics[width=8.6cm]{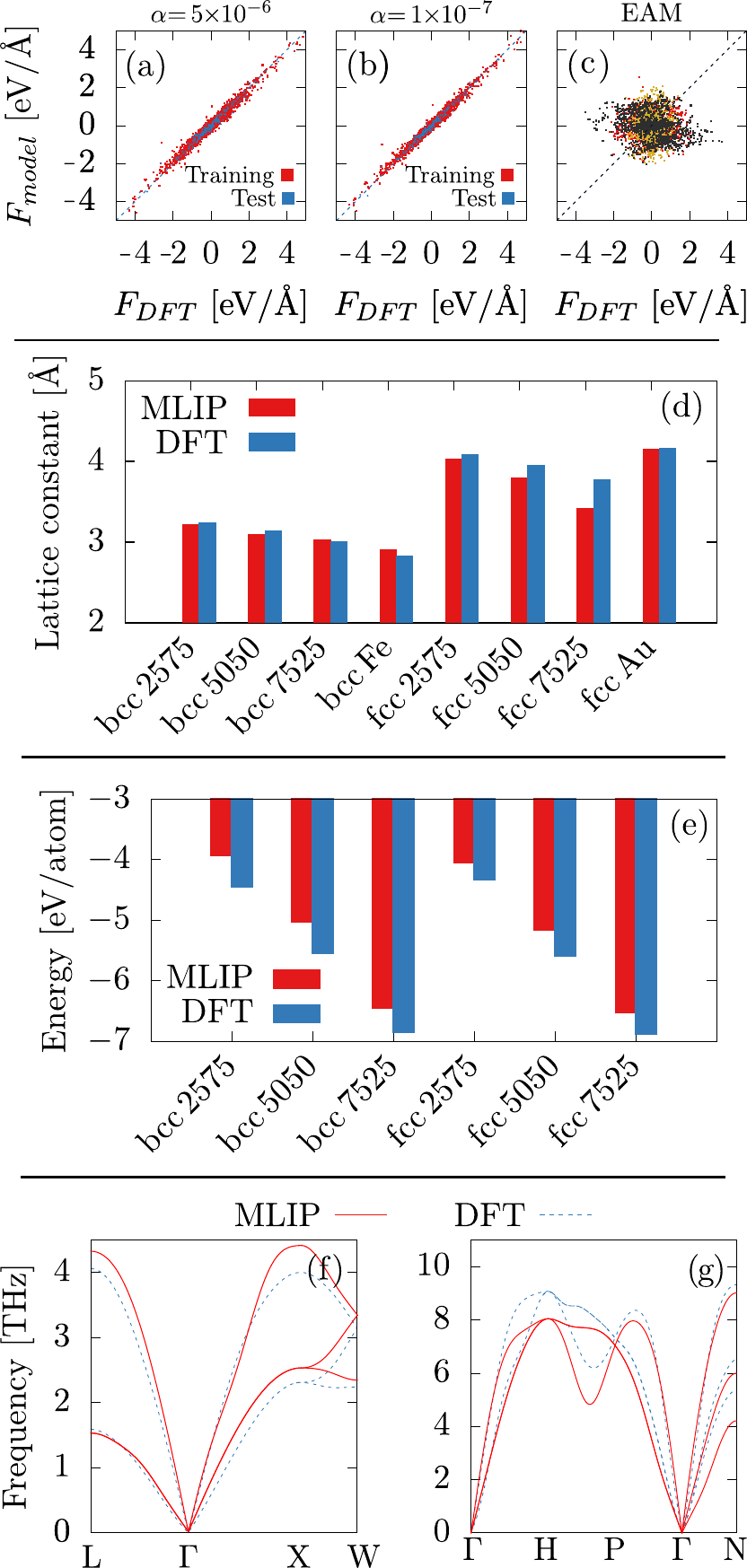}
\caption{(a-c) Correlation plot between DFT forces and forces from MLIP with $\alpha=10^{-7}$ (b) and with $\alpha=5\times10^{-6}$ (c) and from EAM (d).  Blue and red points correspond to results within the training and the test sets. (d) Relative errors obtained on the lattice parameters. (e-f) Phonon dispersion of respectively Gold FCC and Iron BCC. The plain lines correspond to MLIP results while the dotted lines are obtained by DFT calculations.}
\label{Valid}
\end{figure}

Even if the aim of the paper was not to obtain an all-purpose MLIP for gold-iron interactions, we still wish to asses the quality of the obtained potential. Regarding the force accuracy on the learned database, Fig.\,\ref{Valid}(a.b) shows the correlation plot obtained with two different values of $\alpha$. For the RMSE, we obtained in both cases values of $0.14$eV/\AA\,when using the three descriptors at the same time. Such value is on par with most of the currently employed MLIP methods\cite{Nguyen2018Jun} and by comparison, the EAM potential that was recently developed for Au-Fe nanoparticle \cite{Benoit2012Aug} gives a value of $1.4$\,eV/\AA. Our MLIP is thus already a drastic improvement in force evaluation for the studied nanoparticles [See Fig.\ref{Valid}.c]. 

Furthermore, additional simulations with the MLIP were performed to check some properties in the bulk phase although it was not included in the training set. Simulations were carried out using the large-scale molecular dynamics software LAMMPS \citep{plimpton1995} in which the proposed MLIP was implemented. Periodic boundary conditions were employed and the different minimization runs were performed down to a net force of $10^{-6}$\,eV/\AA. We measured eight different lattice constants (pure iron bcc, pure gold fcc, alloys with 25\%, 50\% and 75\% of gold in both bcc and fcc phases). In addition, our fitting did not include any energy. Therefore, to take into account the atomic energy, the MLIP energies are shifted in order to match the cohesive energy of pure iron and gold most stable states ie. bcc and fcc respectively. Then, we also measured the cohesive energy for each alloying structures. Results are compared  in Fig.\,\ref{Valid}.(d,e) to DFT calculations that were previously obtained.\citep{Calvo2017Mar} The errors are lower than $3$\% for lattice spacing and 12\% for cohesive energy which is already satisfying considering that in the database, we employed alloying proportions that are much closer to 50\% and only used nano-crystals. Therefore, being able to reach such small relative errors even for those extreme alloying proportions (25\% and 75\%) and in addition with bulk structures is an encouraging result for our MLIP. 

Finally, we measured phonon dispersion curves using supercell approach implemented in PHONOPY\cite{phonopy} [See Fig.\ref{Valid}(e,f)]. In practice, the dynamical matrix was obtained by moving each symmetrically independent atoms by $0.01$\AA. We used a supercell of size $4\times4\times4$. The agreement between DFT and MLIP curves is not as good as what is usually obtained in MLIP works\cite{Seko2015Aug,Byggmastar2019Oct} but it remains qualitatively satisfying if one considers that our MLIP was not trained on any bulk structures. 

Even if our potential has not been designed to reproduce bulk properties, these results are already very encouraging. Obtaining an accurate MLIP potential that is transferable to any phase and/or structure is a considerable challenge for multi-component systems and would require to carry out additional DFT calculations to build a bigger database including bulk structures but also interfaces. Besides, in order to target a specific application, one should also perform "on-the-fly" optimization of the potential as proposed by Jinnouchi et al.\cite{Jinnouchi2019Jul,Jinnouchi2019Jun}. However this is not the main purpose of this paper, which focuses on shedding light onto the relationship between complexity and transferability in machine learning force fields. It remains that our current MLIP potential may be used as a first step when studying bimetallic Fe-Au nanoparticles.

\section{Discussion and conclusion}

To summarize, this work aims at measuring transferability issues that can occur when using MLIP in untrained structures. To begin, we presented the employed MLIP method that includes a linearized potential and a penalizing regression scheme. Then, we discussed the influence of the descriptor space and showed that although the three different peak functions behave similarly in terms of accuracy and number of non-zero coefficients, the repulsive functions lead to a worst accuracy but with a lower number of non-zero coefficients. Next, we demonstrated that by using the LassoLars algorithm instead of the previously employed linear regression scheme, one can finely tune the complexity of the potential along with its accuracy. This is because the penalizing parameter $\alpha$ allows for turning off most of the initially proposed descriptors thus controlling the overall complexity of the potential. With this ability in hand, we measured transferability issues using three unlearned structures that are qualitatively different from what was considered in the training database. We showed that while the accuracy on the trained structures decrease monotonically as the value of $\alpha$ is decreased, it is not the case for those untrained structures. Indeed, when using only one type of descriptor, it exists an optimal value of $\alpha$ that allows for the best transferability. Finally, we introduced a way to overcome this transferability issue which consists in using  different types of descriptors simultaneously. This again shows an other advantage of using the LassoLars algorithm which is able to actively select the most appropriate descriptors. Finally, we computed some properties of the Fe-Au in bulk and showed that the obtained potential is already qualitatively satisfying. But, before being able to really our potential from practical applications, we plan to improve further it by adding bulk and interface DFT calculations within the database and by implementing "on the fly" learning. As a perspective, the obtained MLIP will be further improved and then employed to study nucleation in core-shell FeAu nanoparticles as observed in experiments.\cite{Langlois2015Aug,Benzo2019Sep,Tymoczko2019,Tymoczko2018Sep,Ponchet2020Aug} Before closing, we wish to discuss two additional points.

\begin{figure}[t!]
\includegraphics[width=8.6cm]{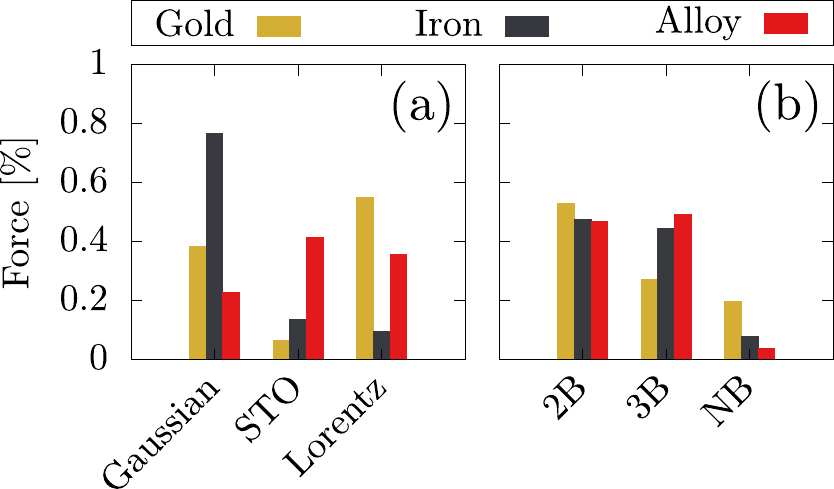}
\caption{Force contributions for each types of (a) descriptors and (b) multi-body components averaged over alloy, pure iron and pure gold nanoparticles.}
\label{Physics}
\end{figure}

First, the very intuitive MLIP expression that was employed here allows us to give some insights on the nature of the interactions. Here, we focus on the combined MLIP where Gaussian, STO and Lorentzian peak were used simultaneously. We can indeed distinguish between each descriptors (Gaussian, STO, Lorentzian). For that purpose, we compute the ratio between the absolute value of the forces given by each descriptor and the sum of the absolute values of the three descriptors and then perform an average over each force components (x,y,z) of all of the atoms within the training database (alloy, pure iron and pure gold nanoparticles). Fig.\,\ref{Physics}.a shows that the preponderant functions are different depending on the considered interactions thus highlting the advantage of using simultaneously the three types of descriptors. More interestingly, the same can be done in order to distinguish between two-body, three-body and N-body contributions [see Fig.\,\ref{Physics}.b]. In our case, the two-body and three-body contributions are more important than the N-body contributions. This may explain why the previously employed EAM potentials that do not possess any explicit angular contributions could not accurately compute the forces.

Moreover, we would like to raise an additional implication of our work in which the transferability issues were measured and connected to the complexity of the potential. Indeed, as previously discussed, when using the LassoLars regression scheme, the complexity of the potential can be adjusted using the penalizing parameter. In the alternative MLIP methods, the same is done by modifying (1) the number of neurons and hidden layers in the case of neural-network potential\cite{Behler2015Aug} and (2) the number of selected configurations after sparsification in the case of gaussian approximation model\cite{Bartok2010Apr,Deringer2017Mar}. For some users of these techniques, the rule of thumb may be to use these adjusting parameters in order to increase the complexity of the potential which necessarily improves the accuracy on the learning database. Yet, our work indicates that transferability issues should be expected by such operation.

%
%Working on gold-iron interactions in nannoparticles, the obtained force RMSE is around $0.1$\,eV/\AA which is much better that the previously employed classical potential\cite{Benoit2012Aug} and is on par with most of the currently employed MLIP methods\cite{Nguyen2018Jun}. A satisfying agreement is also obtained when computing some bulk properties. Before targetting any specific applications, further improvement of the obtained potential should be carried out using for example "on the fly" methods\cite{Jindal2018Nov}. 
%To achieve our primary goal, we then show that the LassoLars regression scheme and . Indeed, increasing the accuracy on the trained configurations and the overall complexity of the potential did not increase monotonically its transferability on untrained systems. This observed transferability issue was furthermore corrected by increasing the diversity of the available descriptors using STO and GTO functions along with the originally employed Gaussian functions. Our findings should encourage to combine different types of descriptors instead of only increasing the number of descriptors. Finally, because a linear regression scheme along with a physically-motivated expression of the potential was employed, we gave some insights regarding the physics of the interactions and showed that in our case, two-body and three-body were the most preponderant.
%
%
%Finally, 

\section*{Appendix}

\subsection{Advantage of using LassoLars against other linear regression scheme}
\begin{figure}[h!]
\includegraphics[width=8.6cm]{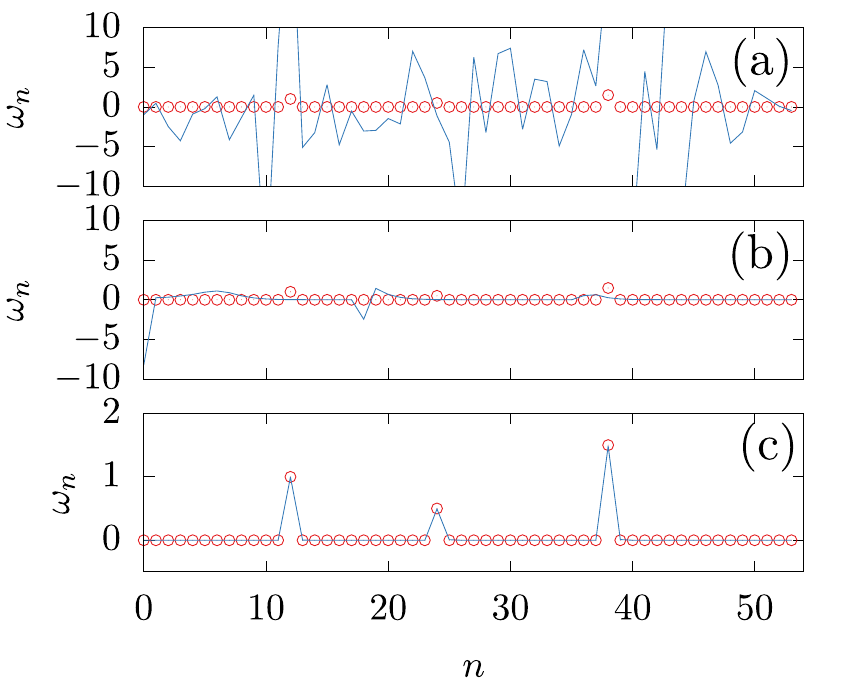}
\caption{Values of the obtained coefficients $\omega_n$ using different linear regression scheme: (a) Ridge, (b) Lasso, (c) LassoLars. The penalty parameter was set to $10^{-5}$. The red points and the blue line correspond respectively to the original interactions and the fitting results.}
\label{LassoLars}
\end{figure}

In this section, we wish to further motivate the choice of our regression scheme by working on a case study where we will compare LassoLars with two commonly employed linear regression schemes naming Ridge and the coordinate descent Lasso. For this test, the database is generated with an equimolar mixture of binary Lennard-Jones particles thus allowing to directly verify the obtained MLIP. With such a binary system, an additional challenge for the fitting algorithm is to distinguish self-species and cross-species interactions. In practice, positions and forces were measured for 50 configurations of 64 atoms in the liquid regime. Regarding the basis of descriptors, only two-body interactions were considered and we used 17 Lennard-Jones functions with different distance parameters including those in the original simulations. All of the four employed methods manage to retrieve a linear combination of Lennard-Jones functions that matches the original interactions. However, it appears that only LassoLars can find the correct coefficients $\omega_n$ setting all coefficients to 0 except those of the original interactions\,[see Fig.\,\ref{LassoLars}]. Such result shows the advantage of using LassoLars instead of the commonly employed linear regression methods.

\subsection{Numerical implementation of the $\Phi$-LassoLars method}

In this section, we give some additional details on the described MLIP. First, obtaining an MLIP using the $\Phi$-LassoLars method consists in two steps. In the first step, a homemade C++ parallelized code using \texttt{OpenMP} was developed to construct a matrix:
\begin{itemize}
\item Each columns of the matrix designates a specific descriptor which can be $2B$, $3B$ or $NB$ for all the functions and considered values for their parameters described in Table 1. 
\item Each rows of the matrix correspond to the force on a given direction, atom and structure. 
\end{itemize}
In the second step, a python code concatenates the obtained matrices for each structures and read the associated DFT forces. The same python code finally employs the LassoLars method as implemented in the \texttt{sklearn} package to obtain the linear coefficients associated to each columns of the matrix.

Then, in order to use the obtained MLIP, the same C++ code is employed to read the obtained coefficients and generate input files for LAMMPS simulations. Those consists on three parts:
\begin{enumerate}
\item For the 2B interactions, we directly add all of the selected linear contributions and generate a table file that can be read by LAMMPS using \texttt{pair\_style table}.
\item For the 3B interactions, we build a homemade routine that is added to LAMMPS and use an personalized input file.
\item For the NB interactions, we use a python code based on  \texttt{atsim.potentials}\cite{atsim.potentials} to generate EAM-like files that can be directly read by LAMMPS.
\end{enumerate}
Finally, the  \texttt{pair\_style hybrid/overlay} is employed to combine all the contributions.

\vspace{0.5cm}

Regarding the computational timing for building the MLIP, a matrix for a structure of $64$ atoms containing all of the functions (2B, 3B, NB for all values in Table 1) for one type of descriptors is obtained in approximately 3 minutes. This process can be parallelized since each structure can be treated independently. Then, after reading all of the input matrices, a LassoLars fitting takes less than 20 seconds when using only one type of functions and less than 2 minutes when using simultaneously three types of functions. Results are obtained on one Intel E5-2650 processor.

%\subsection{Results with a EAM potential fitted with machine-learning potential}
%In this section, we wish to demonstrate that the surprising behavior observed for gold-iron interactions could also be found in a different system. In particular, we worked on pure iron interactions modeled with EAM potentials\cite{Mendelev2003Dec}. Using such classical potentials instead of DFT provides a quick way of creating a force database and also a simpler test that those made of DFT calculations. For the training set, we proceed as in the main article and used a nanoparticle made of $75$ atoms in the BCC phase that is melted to the liquid regime. From such database, several MLIP are trained when varying the constrain parameter $\alpha$. Then, to test the  transferability, we measured the RMSE on a bulk FCC crystal. As we obtained with the gold-iron systems with forces from DFT calculations, it appears that (1) with only Gaussian functions, there is non-monotonic behavior of the RMSE on this untrained structure and (2) this non-monotonic behavior is reduced when adding STO and GTO functions in the descriptor space. 
%
%
%\begin{figure}[h!]
%\includegraphics[width=8.6cm]{Figures/EAM.pdf}
%\caption{Fitting error obtained on pure iron interactions modeled with EAM potentials using only Gaussian functions (dotted line) and adding STO and GTO functions (plain line).}
%\label{LassoLars}
%\end{figure}

$$$$

\section*{Acknowledgement}

JL acknowledges financial support of the Fonds de la Recherche Scientifique - FNRS. Computational resources have been provided by the Consortium des Equipements de Calcul Intensif (CECI), by the F\'ed\'eration Lyonnaise de Mod\'elisation et Sciences Numériques (FLMSN) and by the Regional Computer Center CALMIP in Toulouse (grant n$^{\circ}$ p1141). JL thanks David Tew for having introduced him to the machine-learning potential problem and C. Patrick Royall for supervising the early method development. JL is also grateful to Francesco Turci, Joshua F. Robinson and Atsuto Seko for fruitful discussions. Authors are finally grateful to Abdul-Rahman Allouche and Albert P. Bartók for proofreading the manuscript.


\begin{thebibliography}{66}%
\makeatletter
\providecommand \@ifxundefined [1]{%
 \@ifx{#1\undefined}
}%
\providecommand \@ifnum [1]{%
 \ifnum #1\expandafter \@firstoftwo
 \else \expandafter \@secondoftwo
 \fi
}%
\providecommand \@ifx [1]{%
 \ifx #1\expandafter \@firstoftwo
 \else \expandafter \@secondoftwo
 \fi
}%
\providecommand \natexlab [1]{#1}%
\providecommand \enquote  [1]{``#1''}%
\providecommand \bibnamefont  [1]{#1}%
\providecommand \bibfnamefont [1]{#1}%
\providecommand \citenamefont [1]{#1}%
\providecommand \href@noop [0]{\@secondoftwo}%
\providecommand \href [0]{\begingroup \@sanitize@url \@href}%
\providecommand \@href[1]{\@@startlink{#1}\@@href}%
\providecommand \@@href[1]{\endgroup#1\@@endlink}%
\providecommand \@sanitize@url [0]{\catcode `\\12\catcode `\$12\catcode
  `\&12\catcode `\#12\catcode `\^12\catcode `\_12\catcode `\%12\relax}%
\providecommand \@@startlink[1]{}%
\providecommand \@@endlink[0]{}%
\providecommand \url  [0]{\begingroup\@sanitize@url \@url }%
\providecommand \@url [1]{\endgroup\@href {#1}{\urlprefix }}%
\providecommand \urlprefix  [0]{URL }%
\providecommand \Eprint [0]{\href }%
\providecommand \doibase [0]{http://dx.doi.org/}%
\providecommand \selectlanguage [0]{\@gobble}%
\providecommand \bibinfo  [0]{\@secondoftwo}%
\providecommand \bibfield  [0]{\@secondoftwo}%
\providecommand \translation [1]{[#1]}%
\providecommand \BibitemOpen [0]{}%
\providecommand \bibitemStop [0]{}%
\providecommand \bibitemNoStop [0]{.\EOS\space}%
\providecommand \EOS [0]{\spacefactor3000\relax}%
\providecommand \BibitemShut  [1]{\csname bibitem#1\endcsname}%
\let\auto@bib@innerbib\@empty
%</preamble>
\bibitem [{\citenamefont {Marx}(2009)}]{marx2009ab}%
  \BibitemOpen
  \bibfield  {author} {\bibinfo {author} {\bibfnamefont {D.}~\bibnamefont
  {Marx}},\ }\href@noop {} {\emph {\bibinfo {title} {Ab Initio Molecular
  Dynamics : Basic Theory and Advanced Methods}}}\ (\bibinfo  {publisher}
  {Cambridge University Press},\ \bibinfo {address} {Leiden},\ \bibinfo {year}
  {2009})\BibitemShut {NoStop}%
\bibitem [{\citenamefont {Martin}(2008)}]{martin2008electronic}%
  \BibitemOpen
  \bibfield  {author} {\bibinfo {author} {\bibfnamefont {R.}~\bibnamefont
  {Martin}},\ }\href@noop {} {\emph {\bibinfo {title} {Electronic structure :
  basic theory and practical methods}}}\ (\bibinfo  {publisher} {Cambridge
  University Press},\ \bibinfo {address} {Cambridge, UK New York},\ \bibinfo
  {year} {2008})\BibitemShut {NoStop}%
\bibitem [{\citenamefont {Frenkel}\ and\ \citenamefont
  {Smit}(2002)}]{Frenkel2002}%
  \BibitemOpen
  \bibfield  {author} {\bibinfo {author} {\bibfnamefont {D.}~\bibnamefont
  {Frenkel}}\ and\ \bibinfo {author} {\bibfnamefont {B.}~\bibnamefont {Smit}},\
  }\href {\doibase 10.1016/B978-0-12-267351-1.X5000-7} {\emph {\bibinfo {title}
  {{Understanding Molecular Simulation}}}}\ (\bibinfo  {publisher} {Elsevier,
  Academic Press},\ \bibinfo {year} {2002})\BibitemShut {NoStop}%
\bibitem [{\citenamefont {Stone}(2016)}]{stone2016the}%
  \BibitemOpen
  \bibfield  {author} {\bibinfo {author} {\bibfnamefont {A.~J.}\ \bibnamefont
  {Stone}},\ }\href@noop {} {\emph {\bibinfo {title} {The theory of
  intermolecular forces}}}\ (\bibinfo  {publisher} {Oxford University Press},\
  \bibinfo {address} {Oxford},\ \bibinfo {year} {2016})\BibitemShut {NoStop}%
\bibitem [{\citenamefont {Behler}(2015)}]{Behler2015Aug}%
  \BibitemOpen
  \bibfield  {author} {\bibinfo {author} {\bibfnamefont {J.}~\bibnamefont
  {Behler}},\ }\href {\doibase 10.1002/qua.24890} {\bibfield  {journal}
  {\bibinfo  {journal} {Int. J. Quantum Chem.}\ }\textbf {\bibinfo {volume}
  {115}},\ \bibinfo {pages} {1032} (\bibinfo {year} {2015})}\BibitemShut
  {NoStop}%
\bibitem [{\citenamefont {Bart{\ifmmode\acute{o}\else\'{o}\fi}k}\ and\
  \citenamefont {Cs{\ifmmode\acute{a}\else\'{a}\fi}nyi}(2015)}]{Bartok2015Aug}%
  \BibitemOpen
  \bibfield  {author} {\bibinfo {author} {\bibfnamefont {A.~P.}\ \bibnamefont
  {Bart{\ifmmode\acute{o}\else\'{o}\fi}k}}\ and\ \bibinfo {author}
  {\bibfnamefont {G.}~\bibnamefont {Cs{\ifmmode\acute{a}\else\'{a}\fi}nyi}},\
  }\href {\doibase 10.1002/qua.24927} {\bibfield  {journal} {\bibinfo
  {journal} {Int. J. Quantum Chem.}\ }\textbf {\bibinfo {volume} {115}},\
  \bibinfo {pages} {1051} (\bibinfo {year} {2015})}\BibitemShut {NoStop}%
\bibitem [{\citenamefont {Behler}(2016)}]{Behler2016Nov}%
  \BibitemOpen
  \bibfield  {author} {\bibinfo {author} {\bibfnamefont {J.}~\bibnamefont
  {Behler}},\ }\href {\doibase 10.1063/1.4966192} {\bibfield  {journal}
  {\bibinfo  {journal} {J. Chem. Phys.}\ }\textbf {\bibinfo {volume} {145}},\
  \bibinfo {pages} {170901} (\bibinfo {year} {2016})}\BibitemShut {NoStop}%
\bibitem [{\citenamefont {Behler}\ and\ \citenamefont
  {Parrinello}(2007)}]{Behler2007Apr}%
  \BibitemOpen
  \bibfield  {author} {\bibinfo {author} {\bibfnamefont {J.}~\bibnamefont
  {Behler}}\ and\ \bibinfo {author} {\bibfnamefont {M.}~\bibnamefont
  {Parrinello}},\ }\href {\doibase 10.1103/PhysRevLett.98.146401} {\bibfield
  {journal} {\bibinfo  {journal} {Phys. Rev. Lett.}\ }\textbf {\bibinfo
  {volume} {98}},\ \bibinfo {pages} {146401} (\bibinfo {year}
  {2007})}\BibitemShut {NoStop}%
\bibitem [{\citenamefont {Bart{\ifmmode\acute{o}\else\'{o}\fi}k}\ \emph
  {et~al.}(2010)\citenamefont {Bart{\ifmmode\acute{o}\else\'{o}\fi}k},
  \citenamefont {Payne}, \citenamefont {Kondor},\ and\ \citenamefont
  {Cs{\ifmmode\acute{a}\else\'{a}\fi}nyi}}]{Bartok2010Apr}%
  \BibitemOpen
  \bibfield  {author} {\bibinfo {author} {\bibfnamefont {A.~P.}\ \bibnamefont
  {Bart{\ifmmode\acute{o}\else\'{o}\fi}k}}, \bibinfo {author} {\bibfnamefont
  {M.~C.}\ \bibnamefont {Payne}}, \bibinfo {author} {\bibfnamefont
  {R.}~\bibnamefont {Kondor}}, \ and\ \bibinfo {author} {\bibfnamefont
  {G.}~\bibnamefont {Cs{\ifmmode\acute{a}\else\'{a}\fi}nyi}},\ }\href {\doibase
  10.1103/PhysRevLett.104.136403} {\bibfield  {journal} {\bibinfo  {journal}
  {Phys. Rev. Lett.}\ }\textbf {\bibinfo {volume} {104}},\ \bibinfo {pages}
  {136403} (\bibinfo {year} {2010})}\BibitemShut {NoStop}%
\bibitem [{\citenamefont {Seko}\ \emph {et~al.}(2014)\citenamefont {Seko},
  \citenamefont {Takahashi},\ and\ \citenamefont {Tanaka}}]{Seko2014Jul}%
  \BibitemOpen
  \bibfield  {author} {\bibinfo {author} {\bibfnamefont {A.}~\bibnamefont
  {Seko}}, \bibinfo {author} {\bibfnamefont {A.}~\bibnamefont {Takahashi}}, \
  and\ \bibinfo {author} {\bibfnamefont {I.}~\bibnamefont {Tanaka}},\ }\href
  {\doibase 10.1103/PhysRevB.90.024101} {\bibfield  {journal} {\bibinfo
  {journal} {Phys. Rev. B}\ }\textbf {\bibinfo {volume} {90}},\ \bibinfo
  {pages} {024101} (\bibinfo {year} {2014})}\BibitemShut {NoStop}%
\bibitem [{\citenamefont {Goryaeva}\ \emph {et~al.}(2019)\citenamefont
  {Goryaeva}, \citenamefont {Maillet},\ and\ \citenamefont
  {Marinica}}]{Goryaeva2019Aug}%
  \BibitemOpen
  \bibfield  {author} {\bibinfo {author} {\bibfnamefont {A.~M.}\ \bibnamefont
  {Goryaeva}}, \bibinfo {author} {\bibfnamefont {J.-B.}\ \bibnamefont
  {Maillet}}, \ and\ \bibinfo {author} {\bibfnamefont {M.-C.}\ \bibnamefont
  {Marinica}},\ }\href {\doibase 10.1016/j.commatsci.2019.04.043} {\bibfield
  {journal} {\bibinfo  {journal} {Comput. Mater. Sci.}\ }\textbf {\bibinfo
  {volume} {166}},\ \bibinfo {pages} {200} (\bibinfo {year}
  {2019})}\BibitemShut {NoStop}%
\bibitem [{\citenamefont {Thompson}\ \emph {et~al.}(2015)\citenamefont
  {Thompson}, \citenamefont {Swiler}, \citenamefont {Trott}, \citenamefont
  {Foiles},\ and\ \citenamefont {Tucker}}]{Thompson2015Mar}%
  \BibitemOpen
  \bibfield  {author} {\bibinfo {author} {\bibfnamefont {A.~P.}\ \bibnamefont
  {Thompson}}, \bibinfo {author} {\bibfnamefont {L.~P.}\ \bibnamefont
  {Swiler}}, \bibinfo {author} {\bibfnamefont {C.~R.}\ \bibnamefont {Trott}},
  \bibinfo {author} {\bibfnamefont {S.~M.}\ \bibnamefont {Foiles}}, \ and\
  \bibinfo {author} {\bibfnamefont {G.~J.}\ \bibnamefont {Tucker}},\ }\href
  {\doibase 10.1016/j.jcp.2014.12.018} {\bibfield  {journal} {\bibinfo
  {journal} {J. Comput. Phys.}\ }\textbf {\bibinfo {volume} {285}},\ \bibinfo
  {pages} {316} (\bibinfo {year} {2015})}\BibitemShut {NoStop}%
\bibitem [{\citenamefont {Chmiela}\ \emph {et~al.}(2017)\citenamefont
  {Chmiela}, \citenamefont {Tkatchenko}, \citenamefont {Sauceda}, \citenamefont
  {Poltavsky}, \citenamefont {Sch{\ifmmode\ddot{u}\else\"{u}\fi}tt},\ and\
  \citenamefont {M{\ifmmode\ddot{u}\else\"{u}\fi}ller}}]{Chmiela2017May}%
  \BibitemOpen
  \bibfield  {author} {\bibinfo {author} {\bibfnamefont {S.}~\bibnamefont
  {Chmiela}}, \bibinfo {author} {\bibfnamefont {A.}~\bibnamefont {Tkatchenko}},
  \bibinfo {author} {\bibfnamefont {H.~E.}\ \bibnamefont {Sauceda}}, \bibinfo
  {author} {\bibfnamefont {I.}~\bibnamefont {Poltavsky}}, \bibinfo {author}
  {\bibfnamefont {K.~T.}\ \bibnamefont {Sch{\ifmmode\ddot{u}\else\"{u}\fi}tt}},
  \ and\ \bibinfo {author} {\bibfnamefont {K.-R.}\ \bibnamefont
  {M{\ifmmode\ddot{u}\else\"{u}\fi}ller}},\ }\href {\doibase
  10.1126/sciadv.1603015} {\bibfield  {journal} {\bibinfo  {journal} {Sci.
  Adv.}\ }\textbf {\bibinfo {volume} {3}},\ \bibinfo {pages} {e1603015}
  (\bibinfo {year} {2017})}\BibitemShut {NoStop}%
\bibitem [{\citenamefont {Chmiela}\ \emph {et~al.}(2018)\citenamefont
  {Chmiela}, \citenamefont {Sauceda}, \citenamefont
  {M{\ifmmode\ddot{u}\else\"{u}\fi}ller},\ and\ \citenamefont
  {Tkatchenko}}]{Chmiela2018Sep}%
  \BibitemOpen
  \bibfield  {author} {\bibinfo {author} {\bibfnamefont {S.}~\bibnamefont
  {Chmiela}}, \bibinfo {author} {\bibfnamefont {H.~E.}\ \bibnamefont
  {Sauceda}}, \bibinfo {author} {\bibfnamefont {K.-R.}\ \bibnamefont
  {M{\ifmmode\ddot{u}\else\"{u}\fi}ller}}, \ and\ \bibinfo {author}
  {\bibfnamefont {A.}~\bibnamefont {Tkatchenko}},\ }\href {\doibase
  10.1038/s41467-018-06169-2} {\bibfield  {journal} {\bibinfo  {journal} {Nat.
  Commun.}\ }\textbf {\bibinfo {volume} {9}},\ \bibinfo {pages} {1} (\bibinfo
  {year} {2018})}\BibitemShut {NoStop}%
\bibitem [{\citenamefont {Shapeev}(2016)}]{Shapeev2016Sep}%
  \BibitemOpen
  \bibfield  {author} {\bibinfo {author} {\bibfnamefont {A.~V.}\ \bibnamefont
  {Shapeev}},\ }\href {https://epubs.siam.org/doi/10.1137/15M1054183}
  {\bibfield  {journal} {\bibinfo  {journal} {Multiscale Model. Simul.}\ }
  (\bibinfo {year} {2016})}\BibitemShut {NoStop}%
\bibitem [{\citenamefont {Novoselov}\ \emph {et~al.}(2019)\citenamefont
  {Novoselov}, \citenamefont {Yanilkin}, \citenamefont {Shapeev},\ and\
  \citenamefont {Podryabinkin}}]{Novoselov2019Jun}%
  \BibitemOpen
  \bibfield  {author} {\bibinfo {author} {\bibfnamefont {I.~I.}\ \bibnamefont
  {Novoselov}}, \bibinfo {author} {\bibfnamefont {A.~V.}\ \bibnamefont
  {Yanilkin}}, \bibinfo {author} {\bibfnamefont {A.~V.}\ \bibnamefont
  {Shapeev}}, \ and\ \bibinfo {author} {\bibfnamefont {E.~V.}\ \bibnamefont
  {Podryabinkin}},\ }\href {\doibase 10.1016/j.commatsci.2019.03.049}
  {\bibfield  {journal} {\bibinfo  {journal} {Comput. Mater. Sci.}\ }\textbf
  {\bibinfo {volume} {164}},\ \bibinfo {pages} {46} (\bibinfo {year}
  {2019})}\BibitemShut {NoStop}%
\bibitem [{\citenamefont {Seko}\ \emph {et~al.}(2015)\citenamefont {Seko},
  \citenamefont {Takahashi},\ and\ \citenamefont {Tanaka}}]{Seko2015Aug}%
  \BibitemOpen
  \bibfield  {author} {\bibinfo {author} {\bibfnamefont {A.}~\bibnamefont
  {Seko}}, \bibinfo {author} {\bibfnamefont {A.}~\bibnamefont {Takahashi}}, \
  and\ \bibinfo {author} {\bibfnamefont {I.}~\bibnamefont {Tanaka}},\ }\href
  {\doibase 10.1103/PhysRevB.92.054113} {\bibfield  {journal} {\bibinfo
  {journal} {Phys. Rev. B}\ }\textbf {\bibinfo {volume} {92}},\ \bibinfo
  {pages} {054113} (\bibinfo {year} {2015})}\BibitemShut {NoStop}%
\bibitem [{\citenamefont {Takahashi}\ \emph {et~al.}(2018)\citenamefont
  {Takahashi}, \citenamefont {Seko},\ and\ \citenamefont
  {Tanaka}}]{Takahashi2018Jun}%
  \BibitemOpen
  \bibfield  {author} {\bibinfo {author} {\bibfnamefont {A.}~\bibnamefont
  {Takahashi}}, \bibinfo {author} {\bibfnamefont {A.}~\bibnamefont {Seko}}, \
  and\ \bibinfo {author} {\bibfnamefont {I.}~\bibnamefont {Tanaka}},\ }\href
  {\doibase 10.1063/1.5027283} {\bibfield  {journal} {\bibinfo  {journal} {J.
  Chem. Phys.}\ }\textbf {\bibinfo {volume} {148}},\ \bibinfo {pages} {234106}
  (\bibinfo {year} {2018})}\BibitemShut {NoStop}%
\bibitem [{\citenamefont {Zeni}\ \emph {et~al.}(2018)\citenamefont {Zeni},
  \citenamefont {Rossi}, \citenamefont {Glielmo}, \citenamefont {Fekete},
  \citenamefont {Gaston}, \citenamefont {Baletto},\ and\ \citenamefont
  {De~Vita}}]{Zeni2018Jun}%
  \BibitemOpen
  \bibfield  {author} {\bibinfo {author} {\bibfnamefont {C.}~\bibnamefont
  {Zeni}}, \bibinfo {author} {\bibfnamefont {K.}~\bibnamefont {Rossi}},
  \bibinfo {author} {\bibfnamefont {A.}~\bibnamefont {Glielmo}}, \bibinfo
  {author} {\bibfnamefont {{\ifmmode\acute{A}\else\'{A}\fi}.}~\bibnamefont
  {Fekete}}, \bibinfo {author} {\bibfnamefont {N.}~\bibnamefont {Gaston}},
  \bibinfo {author} {\bibfnamefont {F.}~\bibnamefont {Baletto}}, \ and\
  \bibinfo {author} {\bibfnamefont {A.}~\bibnamefont {De~Vita}},\ }\href
  {\doibase 10.1063/1.5024558} {\bibfield  {journal} {\bibinfo  {journal} {J.
  Chem. Phys.}\ }\textbf {\bibinfo {volume} {148}},\ \bibinfo {pages} {241739}
  (\bibinfo {year} {2018})}\BibitemShut {NoStop}%
\bibitem [{\citenamefont {Botu}\ \emph {et~al.}(2017)\citenamefont {Botu},
  \citenamefont {Batra}, \citenamefont {Chapman},\ and\ \citenamefont
  {Ramprasad}}]{Botu2017Jan}%
  \BibitemOpen
  \bibfield  {author} {\bibinfo {author} {\bibfnamefont {V.}~\bibnamefont
  {Botu}}, \bibinfo {author} {\bibfnamefont {R.}~\bibnamefont {Batra}},
  \bibinfo {author} {\bibfnamefont {J.}~\bibnamefont {Chapman}}, \ and\
  \bibinfo {author} {\bibfnamefont {R.}~\bibnamefont {Ramprasad}},\ }\href
  {\doibase 10.1021/acs.jpcc.6b10908} {\bibfield  {journal} {\bibinfo
  {journal} {J. Phys. Chem. C}\ }\textbf {\bibinfo {volume} {121}},\ \bibinfo
  {pages} {511} (\bibinfo {year} {2017})}\BibitemShut {NoStop}%
\bibitem [{\citenamefont {Bereau}\ \emph {et~al.}(2018)\citenamefont {Bereau},
  \citenamefont {DiStasio}, \citenamefont {Tkatchenko},\ and\ \citenamefont
  {von Lilienfeld}}]{Bereau2018Jun}%
  \BibitemOpen
  \bibfield  {author} {\bibinfo {author} {\bibfnamefont {T.}~\bibnamefont
  {Bereau}}, \bibinfo {author} {\bibfnamefont {R.~A.}\ \bibnamefont
  {DiStasio}}, \bibinfo {author} {\bibfnamefont {A.}~\bibnamefont
  {Tkatchenko}}, \ and\ \bibinfo {author} {\bibfnamefont {O.~A.}\ \bibnamefont
  {von Lilienfeld}},\ }\href {\doibase
  10.1063/1.5009502@jcp.2018.DETC2018.issue-1} {\bibfield  {journal} {\bibinfo
  {journal} {J. Chem. Phys.}\ }\textbf {\bibinfo {volume} {DETC2018}},\
  \bibinfo {pages} {241706} (\bibinfo {year} {2018})}\BibitemShut {NoStop}%
\bibitem [{\citenamefont {Sauceda}\ \emph {et~al.}(2019)\citenamefont
  {Sauceda}, \citenamefont {Chmiela}, \citenamefont {Poltavsky}, \citenamefont
  {M{\ifmmode\ddot{u}\else\"{u}\fi}ller},\ and\ \citenamefont
  {Tkatchenko}}]{Sauceda2019Mar}%
  \BibitemOpen
  \bibfield  {author} {\bibinfo {author} {\bibfnamefont {H.~E.}\ \bibnamefont
  {Sauceda}}, \bibinfo {author} {\bibfnamefont {S.}~\bibnamefont {Chmiela}},
  \bibinfo {author} {\bibfnamefont {I.}~\bibnamefont {Poltavsky}}, \bibinfo
  {author} {\bibfnamefont {K.-R.}\ \bibnamefont
  {M{\ifmmode\ddot{u}\else\"{u}\fi}ller}}, \ and\ \bibinfo {author}
  {\bibfnamefont {A.}~\bibnamefont {Tkatchenko}},\ }\href {\doibase
  10.1063/1.5078687} {\bibfield  {journal} {\bibinfo  {journal} {J. Chem.
  Phys.}\ }\textbf {\bibinfo {volume} {150}},\ \bibinfo {pages} {114102}
  (\bibinfo {year} {2019})}\BibitemShut {NoStop}%
\bibitem [{\citenamefont {Bart{\ifmmode\acute{o}\else\'{o}\fi}k}\ \emph
  {et~al.}(2017)\citenamefont {Bart{\ifmmode\acute{o}\else\'{o}\fi}k},
  \citenamefont {De}, \citenamefont {Poelking}, \citenamefont {Bernstein},
  \citenamefont {Kermode}, \citenamefont
  {Cs{\ifmmode\acute{a}\else\'{a}\fi}nyi},\ and\ \citenamefont
  {Ceriotti}}]{Bartok2017Dec}%
  \BibitemOpen
  \bibfield  {author} {\bibinfo {author} {\bibfnamefont {A.~P.}\ \bibnamefont
  {Bart{\ifmmode\acute{o}\else\'{o}\fi}k}}, \bibinfo {author} {\bibfnamefont
  {S.}~\bibnamefont {De}}, \bibinfo {author} {\bibfnamefont {C.}~\bibnamefont
  {Poelking}}, \bibinfo {author} {\bibfnamefont {N.}~\bibnamefont {Bernstein}},
  \bibinfo {author} {\bibfnamefont {J.~R.}\ \bibnamefont {Kermode}}, \bibinfo
  {author} {\bibfnamefont {G.}~\bibnamefont
  {Cs{\ifmmode\acute{a}\else\'{a}\fi}nyi}}, \ and\ \bibinfo {author}
  {\bibfnamefont {M.}~\bibnamefont {Ceriotti}},\ }\href {\doibase
  10.1126/sciadv.1701816} {\bibfield  {journal} {\bibinfo  {journal} {Sci.
  Adv.}\ }\textbf {\bibinfo {volume} {3}},\ \bibinfo {pages} {e1701816}
  (\bibinfo {year} {2017})}\BibitemShut {NoStop}%
\bibitem [{\citenamefont {Veit}\ \emph {et~al.}(2019)\citenamefont {Veit},
  \citenamefont {Jain}, \citenamefont {Bonakala}, \citenamefont {Rudra},
  \citenamefont {Hohl},\ and\ \citenamefont
  {Cs{\ifmmode\acute{a}\else\'{a}\fi}nyi}}]{Veit2019Apr}%
  \BibitemOpen
  \bibfield  {author} {\bibinfo {author} {\bibfnamefont {M.}~\bibnamefont
  {Veit}}, \bibinfo {author} {\bibfnamefont {S.~K.}\ \bibnamefont {Jain}},
  \bibinfo {author} {\bibfnamefont {S.}~\bibnamefont {Bonakala}}, \bibinfo
  {author} {\bibfnamefont {I.}~\bibnamefont {Rudra}}, \bibinfo {author}
  {\bibfnamefont {D.}~\bibnamefont {Hohl}}, \ and\ \bibinfo {author}
  {\bibfnamefont {G.}~\bibnamefont {Cs{\ifmmode\acute{a}\else\'{a}\fi}nyi}},\
  }\href {\doibase 10.1021/acs.jctc.8b01242} {\bibfield  {journal} {\bibinfo
  {journal} {J. Chem. Theory Comput.}\ }\textbf {\bibinfo {volume} {15}},\
  \bibinfo {pages} {2574} (\bibinfo {year} {2019})}\BibitemShut {NoStop}%
\bibitem [{\citenamefont {Artrith}\ \emph {et~al.}(2011)\citenamefont
  {Artrith}, \citenamefont {Morawietz},\ and\ \citenamefont
  {Behler}}]{Artrith2011Apr}%
  \BibitemOpen
  \bibfield  {author} {\bibinfo {author} {\bibfnamefont {N.}~\bibnamefont
  {Artrith}}, \bibinfo {author} {\bibfnamefont {T.}~\bibnamefont {Morawietz}},
  \ and\ \bibinfo {author} {\bibfnamefont {J.}~\bibnamefont {Behler}},\ }\href
  {\doibase 10.1103/PhysRevB.83.153101} {\bibfield  {journal} {\bibinfo
  {journal} {Phys. Rev. B}\ }\textbf {\bibinfo {volume} {83}},\ \bibinfo
  {pages} {153101} (\bibinfo {year} {2011})}\BibitemShut {NoStop}%
\bibitem [{\citenamefont {Quaranta}\ \emph {et~al.}(2017)\citenamefont
  {Quaranta}, \citenamefont {Hellstr{\ifmmode\ddot{o}\else\"{o}\fi}m},\ and\
  \citenamefont {Behler}}]{Quaranta2017Apr}%
  \BibitemOpen
  \bibfield  {author} {\bibinfo {author} {\bibfnamefont {V.}~\bibnamefont
  {Quaranta}}, \bibinfo {author} {\bibfnamefont {M.}~\bibnamefont
  {Hellstr{\ifmmode\ddot{o}\else\"{o}\fi}m}}, \ and\ \bibinfo {author}
  {\bibfnamefont {J.}~\bibnamefont {Behler}},\ }\href {\doibase
  10.1021/acs.jpclett.7b00358} {\bibfield  {journal} {\bibinfo  {journal} {J.
  Phys. Chem. Lett.}\ }\textbf {\bibinfo {volume} {8}},\ \bibinfo {pages}
  {1476} (\bibinfo {year} {2017})}\BibitemShut {NoStop}%
\bibitem [{\citenamefont {Nguyen}\ \emph {et~al.}(2018)\citenamefont {Nguyen},
  \citenamefont {Sz{\ifmmode\acute{e}\else\'{e}\fi}kely}, \citenamefont
  {Imbalzano}, \citenamefont {Behler}, \citenamefont
  {Cs{\ifmmode\acute{a}\else\'{a}\fi}nyi}, \citenamefont {Ceriotti},
  \citenamefont {G{\ifmmode\ddot{o}\else\"{o}\fi}tz},\ and\ \citenamefont
  {Paesani}}]{Nguyen2018Jun}%
  \BibitemOpen
  \bibfield  {author} {\bibinfo {author} {\bibfnamefont {T.~T.}\ \bibnamefont
  {Nguyen}}, \bibinfo {author} {\bibfnamefont {E.}~\bibnamefont
  {Sz{\ifmmode\acute{e}\else\'{e}\fi}kely}}, \bibinfo {author} {\bibfnamefont
  {G.}~\bibnamefont {Imbalzano}}, \bibinfo {author} {\bibfnamefont
  {J.}~\bibnamefont {Behler}}, \bibinfo {author} {\bibfnamefont
  {G.}~\bibnamefont {Cs{\ifmmode\acute{a}\else\'{a}\fi}nyi}}, \bibinfo {author}
  {\bibfnamefont {M.}~\bibnamefont {Ceriotti}}, \bibinfo {author}
  {\bibfnamefont {A.~W.}\ \bibnamefont {G{\ifmmode\ddot{o}\else\"{o}\fi}tz}}, \
  and\ \bibinfo {author} {\bibfnamefont {F.}~\bibnamefont {Paesani}},\ }\href
  {\doibase 10.1063/1.5024577@jcp.2018.DETC2018.issue-1} {\bibfield  {journal}
  {\bibinfo  {journal} {J. Chem. Phys.}\ }\textbf {\bibinfo {volume}
  {DETC2018}},\ \bibinfo {pages} {241725} (\bibinfo {year} {2018})}\BibitemShut
  {NoStop}%
\bibitem [{\citenamefont {Bart{\ifmmode\acute{o}\else\'{o}\fi}k}\ \emph
  {et~al.}(2013)\citenamefont {Bart{\ifmmode\acute{o}\else\'{o}\fi}k},
  \citenamefont {Gillan}, \citenamefont {Manby},\ and\ \citenamefont
  {Cs{\ifmmode\acute{a}\else\'{a}\fi}nyi}}]{Bartok2013Aug}%
  \BibitemOpen
  \bibfield  {author} {\bibinfo {author} {\bibfnamefont {A.~P.}\ \bibnamefont
  {Bart{\ifmmode\acute{o}\else\'{o}\fi}k}}, \bibinfo {author} {\bibfnamefont
  {M.~J.}\ \bibnamefont {Gillan}}, \bibinfo {author} {\bibfnamefont {F.~R.}\
  \bibnamefont {Manby}}, \ and\ \bibinfo {author} {\bibfnamefont
  {G.}~\bibnamefont {Cs{\ifmmode\acute{a}\else\'{a}\fi}nyi}},\ }\href {\doibase
  10.1103/PhysRevB.88.054104} {\bibfield  {journal} {\bibinfo  {journal} {Phys.
  Rev. B}\ }\textbf {\bibinfo {volume} {88}},\ \bibinfo {pages} {054104}
  (\bibinfo {year} {2013})}\BibitemShut {NoStop}%
\bibitem [{\citenamefont {Morawietz}\ \emph {et~al.}(2012)\citenamefont
  {Morawietz}, \citenamefont {Sharma},\ and\ \citenamefont
  {Behler}}]{Morawietz2012Feb}%
  \BibitemOpen
  \bibfield  {author} {\bibinfo {author} {\bibfnamefont {T.}~\bibnamefont
  {Morawietz}}, \bibinfo {author} {\bibfnamefont {V.}~\bibnamefont {Sharma}}, \
  and\ \bibinfo {author} {\bibfnamefont {J.}~\bibnamefont {Behler}},\ }\href
  {\doibase 10.1063/1.3682557} {\bibfield  {journal} {\bibinfo  {journal} {J.
  Chem. Phys.}\ }\textbf {\bibinfo {volume} {136}},\ \bibinfo {pages} {064103}
  (\bibinfo {year} {2012})}\BibitemShut {NoStop}%
\bibitem [{\citenamefont {Natarajan}\ and\ \citenamefont
  {Behler}(2016)}]{Natarajan2016Oct}%
  \BibitemOpen
  \bibfield  {author} {\bibinfo {author} {\bibfnamefont {S.~K.}\ \bibnamefont
  {Natarajan}}\ and\ \bibinfo {author} {\bibfnamefont {J.}~\bibnamefont
  {Behler}},\ }\href {\doibase 10.1039/C6CP05711J} {\bibfield  {journal}
  {\bibinfo  {journal} {Phys. Chem. Chem. Phys.}\ }\textbf {\bibinfo {volume}
  {18}},\ \bibinfo {pages} {28704} (\bibinfo {year} {2016})}\BibitemShut
  {NoStop}%
\bibitem [{\citenamefont {Morawietz}\ and\ \citenamefont
  {Behler}(2013)}]{Morawietz2013Aug}%
  \BibitemOpen
  \bibfield  {author} {\bibinfo {author} {\bibfnamefont {T.}~\bibnamefont
  {Morawietz}}\ and\ \bibinfo {author} {\bibfnamefont {J.}~\bibnamefont
  {Behler}},\ }\href {\doibase 10.1021/jp401225b} {\bibfield  {journal}
  {\bibinfo  {journal} {J. Phys. Chem. A}\ }\textbf {\bibinfo {volume} {117}},\
  \bibinfo {pages} {7356} (\bibinfo {year} {2013})}\BibitemShut {NoStop}%
\bibitem [{\citenamefont {Deringer}\ and\ \citenamefont
  {Cs{\ifmmode\acute{a}\else\'{a}\fi}nyi}(2017)}]{Deringer2017Mar}%
  \BibitemOpen
  \bibfield  {author} {\bibinfo {author} {\bibfnamefont {V.~L.}\ \bibnamefont
  {Deringer}}\ and\ \bibinfo {author} {\bibfnamefont {G.}~\bibnamefont
  {Cs{\ifmmode\acute{a}\else\'{a}\fi}nyi}},\ }\href {\doibase
  10.1103/PhysRevB.95.094203} {\bibfield  {journal} {\bibinfo  {journal} {Phys.
  Rev. B}\ }\textbf {\bibinfo {volume} {95}},\ \bibinfo {pages} {094203}
  (\bibinfo {year} {2017})}\BibitemShut {NoStop}%
\bibitem [{\citenamefont {Bart{\ifmmode\acute{o}\else\'{o}\fi}k}\ \emph
  {et~al.}(2018)\citenamefont {Bart{\ifmmode\acute{o}\else\'{o}\fi}k},
  \citenamefont {Kermode}, \citenamefont {Bernstein},\ and\ \citenamefont
  {Cs{\ifmmode\acute{a}\else\'{a}\fi}nyi}}]{Bartok2018Dec}%
  \BibitemOpen
  \bibfield  {author} {\bibinfo {author} {\bibfnamefont {A.~P.}\ \bibnamefont
  {Bart{\ifmmode\acute{o}\else\'{o}\fi}k}}, \bibinfo {author} {\bibfnamefont
  {J.}~\bibnamefont {Kermode}}, \bibinfo {author} {\bibfnamefont
  {N.}~\bibnamefont {Bernstein}}, \ and\ \bibinfo {author} {\bibfnamefont
  {G.}~\bibnamefont {Cs{\ifmmode\acute{a}\else\'{a}\fi}nyi}},\ }\href {\doibase
  10.1103/PhysRevX.8.041048} {\bibfield  {journal} {\bibinfo  {journal} {Phys.
  Rev. X}\ }\textbf {\bibinfo {volume} {8}},\ \bibinfo {pages} {041048}
  (\bibinfo {year} {2018})}\BibitemShut {NoStop}%
\bibitem [{\citenamefont {Caro}\ \emph {et~al.}(2018)\citenamefont {Caro},
  \citenamefont {Deringer}, \citenamefont {Koskinen}, \citenamefont {Laurila},\
  and\ \citenamefont {Cs{\ifmmode\acute{a}\else\'{a}\fi}nyi}}]{Caro2018Apr}%
  \BibitemOpen
  \bibfield  {author} {\bibinfo {author} {\bibfnamefont {M.~A.}\ \bibnamefont
  {Caro}}, \bibinfo {author} {\bibfnamefont {V.~L.}\ \bibnamefont {Deringer}},
  \bibinfo {author} {\bibfnamefont {J.}~\bibnamefont {Koskinen}}, \bibinfo
  {author} {\bibfnamefont {T.}~\bibnamefont {Laurila}}, \ and\ \bibinfo
  {author} {\bibfnamefont {G.}~\bibnamefont
  {Cs{\ifmmode\acute{a}\else\'{a}\fi}nyi}},\ }\href {\doibase
  10.1103/PhysRevLett.120.166101} {\bibfield  {journal} {\bibinfo  {journal}
  {Phys. Rev. Lett.}\ }\textbf {\bibinfo {volume} {120}},\ \bibinfo {pages}
  {166101} (\bibinfo {year} {2018})}\BibitemShut {NoStop}%
\bibitem [{\citenamefont {Deringer}\ \emph
  {et~al.}(2018{\natexlab{a}})\citenamefont {Deringer}, \citenamefont
  {Bernstein}, \citenamefont {Bart{\ifmmode\acute{o}\else\'{o}\fi}k},
  \citenamefont {Cliffe}, \citenamefont {Kerber}, \citenamefont {Marbella},
  \citenamefont {Grey}, \citenamefont {Elliott},\ and\ \citenamefont
  {Cs{\ifmmode\acute{a}\else\'{a}\fi}nyi}}]{Deringer2018Jun}%
  \BibitemOpen
  \bibfield  {author} {\bibinfo {author} {\bibfnamefont {V.~L.}\ \bibnamefont
  {Deringer}}, \bibinfo {author} {\bibfnamefont {N.}~\bibnamefont {Bernstein}},
  \bibinfo {author} {\bibfnamefont {A.~P.}\ \bibnamefont
  {Bart{\ifmmode\acute{o}\else\'{o}\fi}k}}, \bibinfo {author} {\bibfnamefont
  {M.~J.}\ \bibnamefont {Cliffe}}, \bibinfo {author} {\bibfnamefont {R.~N.}\
  \bibnamefont {Kerber}}, \bibinfo {author} {\bibfnamefont {L.~E.}\
  \bibnamefont {Marbella}}, \bibinfo {author} {\bibfnamefont {C.~P.}\
  \bibnamefont {Grey}}, \bibinfo {author} {\bibfnamefont {S.~R.}\ \bibnamefont
  {Elliott}}, \ and\ \bibinfo {author} {\bibfnamefont {G.}~\bibnamefont
  {Cs{\ifmmode\acute{a}\else\'{a}\fi}nyi}},\ }\href {\doibase
  10.1021/acs.jpclett.8b00902} {\bibfield  {journal} {\bibinfo  {journal} {J.
  Phys. Chem. Lett.}\ }\textbf {\bibinfo {volume} {9}},\ \bibinfo {pages}
  {2879} (\bibinfo {year} {2018}{\natexlab{a}})}\BibitemShut {NoStop}%
\bibitem [{\citenamefont {Deringer}\ \emph
  {et~al.}(2018{\natexlab{b}})\citenamefont {Deringer}, \citenamefont {Caro},
  \citenamefont {Jana}, \citenamefont {Aarva}, \citenamefont {Elliott},
  \citenamefont {Laurila}, \citenamefont
  {Cs{\ifmmode\acute{a}\else\'{a}\fi}nyi},\ and\ \citenamefont
  {Pastewka}}]{Deringer2018Nov}%
  \BibitemOpen
  \bibfield  {author} {\bibinfo {author} {\bibfnamefont {V.~L.}\ \bibnamefont
  {Deringer}}, \bibinfo {author} {\bibfnamefont {M.~A.}\ \bibnamefont {Caro}},
  \bibinfo {author} {\bibfnamefont {R.}~\bibnamefont {Jana}}, \bibinfo {author}
  {\bibfnamefont {A.}~\bibnamefont {Aarva}}, \bibinfo {author} {\bibfnamefont
  {S.~R.}\ \bibnamefont {Elliott}}, \bibinfo {author} {\bibfnamefont
  {T.}~\bibnamefont {Laurila}}, \bibinfo {author} {\bibfnamefont
  {G.}~\bibnamefont {Cs{\ifmmode\acute{a}\else\'{a}\fi}nyi}}, \ and\ \bibinfo
  {author} {\bibfnamefont {L.}~\bibnamefont {Pastewka}},\ }\href {\doibase
  10.1021/acs.chemmater.8b02410} {\bibfield  {journal} {\bibinfo  {journal}
  {Chem. Mater.}\ }\textbf {\bibinfo {volume} {30}},\ \bibinfo {pages} {7438}
  (\bibinfo {year} {2018}{\natexlab{b}})}\BibitemShut {NoStop}%
\bibitem [{\citenamefont {Sosso}\ \emph {et~al.}(2018)\citenamefont {Sosso},
  \citenamefont {Deringer}, \citenamefont {Elliott},\ and\ \citenamefont
  {Cs{\ifmmode\acute{a}\else\'{a}\fi}nyi}}]{Sosso2018Jul}%
  \BibitemOpen
  \bibfield  {author} {\bibinfo {author} {\bibfnamefont {G.~C.}\ \bibnamefont
  {Sosso}}, \bibinfo {author} {\bibfnamefont {V.~L.}\ \bibnamefont {Deringer}},
  \bibinfo {author} {\bibfnamefont {S.~R.}\ \bibnamefont {Elliott}}, \ and\
  \bibinfo {author} {\bibfnamefont {G.}~\bibnamefont
  {Cs{\ifmmode\acute{a}\else\'{a}\fi}nyi}},\ }\href {\doibase
  10.1080/08927022.2018.1447107} {\bibfield  {journal} {\bibinfo  {journal}
  {Mol. Simul.}\ }\textbf {\bibinfo {volume} {44}},\ \bibinfo {pages} {866}
  (\bibinfo {year} {2018})}\BibitemShut {NoStop}%
\bibitem [{\citenamefont {Jinnouchi}\ \emph
  {et~al.}(2019{\natexlab{a}})\citenamefont {Jinnouchi}, \citenamefont
  {Lahnsteiner}, \citenamefont {Karsai}, \citenamefont {Kresse},\ and\
  \citenamefont {Bokdam}}]{Jinnouchi2019Jun}%
  \BibitemOpen
  \bibfield  {author} {\bibinfo {author} {\bibfnamefont {R.}~\bibnamefont
  {Jinnouchi}}, \bibinfo {author} {\bibfnamefont {J.}~\bibnamefont
  {Lahnsteiner}}, \bibinfo {author} {\bibfnamefont {F.}~\bibnamefont {Karsai}},
  \bibinfo {author} {\bibfnamefont {G.}~\bibnamefont {Kresse}}, \ and\ \bibinfo
  {author} {\bibfnamefont {M.}~\bibnamefont {Bokdam}},\ }\href {\doibase
  10.1103/PhysRevLett.122.225701} {\bibfield  {journal} {\bibinfo  {journal}
  {Phys. Rev. Lett.}\ }\textbf {\bibinfo {volume} {122}},\ \bibinfo {pages}
  {225701} (\bibinfo {year} {2019}{\natexlab{a}})}\BibitemShut {NoStop}%
\bibitem [{\citenamefont {Jinnouchi}\ \emph
  {et~al.}(2019{\natexlab{b}})\citenamefont {Jinnouchi}, \citenamefont
  {Karsai},\ and\ \citenamefont {Kresse}}]{Jinnouchi2019Jul}%
  \BibitemOpen
  \bibfield  {author} {\bibinfo {author} {\bibfnamefont {R.}~\bibnamefont
  {Jinnouchi}}, \bibinfo {author} {\bibfnamefont {F.}~\bibnamefont {Karsai}}, \
  and\ \bibinfo {author} {\bibfnamefont {G.}~\bibnamefont {Kresse}},\ }\href
  {\doibase 10.1103/PhysRevB.100.014105} {\bibfield  {journal} {\bibinfo
  {journal} {Phys. Rev. B}\ }\textbf {\bibinfo {volume} {100}},\ \bibinfo
  {pages} {014105} (\bibinfo {year} {2019}{\natexlab{b}})}\BibitemShut
  {NoStop}%
\bibitem [{\citenamefont {Li}\ \emph {et~al.}(2015)\citenamefont {Li},
  \citenamefont {Kermode},\ and\ \citenamefont {De~Vita}}]{Li2015Mar}%
  \BibitemOpen
  \bibfield  {author} {\bibinfo {author} {\bibfnamefont {Z.}~\bibnamefont
  {Li}}, \bibinfo {author} {\bibfnamefont {J.~R.}\ \bibnamefont {Kermode}}, \
  and\ \bibinfo {author} {\bibfnamefont {A.}~\bibnamefont {De~Vita}},\ }\href
  {\doibase 10.1103/PhysRevLett.114.096405} {\bibfield  {journal} {\bibinfo
  {journal} {Phys. Rev. Lett.}\ }\textbf {\bibinfo {volume} {114}},\ \bibinfo
  {pages} {096405} (\bibinfo {year} {2015})}\BibitemShut {NoStop}%
\bibitem [{\citenamefont {Langlois}\ \emph {et~al.}(2015)\citenamefont
  {Langlois}, \citenamefont {Benzo}, \citenamefont {Arenal}, \citenamefont
  {Benoit}, \citenamefont {Nicolai}, \citenamefont {Combe}, \citenamefont
  {Ponchet},\ and\ \citenamefont {Casanove}}]{Langlois2015Aug}%
  \BibitemOpen
  \bibfield  {author} {\bibinfo {author} {\bibfnamefont {C.}~\bibnamefont
  {Langlois}}, \bibinfo {author} {\bibfnamefont {P.}~\bibnamefont {Benzo}},
  \bibinfo {author} {\bibfnamefont {R.}~\bibnamefont {Arenal}}, \bibinfo
  {author} {\bibfnamefont {M.}~\bibnamefont {Benoit}}, \bibinfo {author}
  {\bibfnamefont {J.}~\bibnamefont {Nicolai}}, \bibinfo {author} {\bibfnamefont
  {N.}~\bibnamefont {Combe}}, \bibinfo {author} {\bibfnamefont
  {A.}~\bibnamefont {Ponchet}}, \ and\ \bibinfo {author} {\bibfnamefont
  {M.~J.}\ \bibnamefont {Casanove}},\ }\href {\doibase
  10.1021/acs.nanolett.5b02273} {\bibfield  {journal} {\bibinfo  {journal}
  {Nano Lett.}\ }\textbf {\bibinfo {volume} {15}},\ \bibinfo {pages} {5075}
  (\bibinfo {year} {2015})}\BibitemShut {NoStop}%
\bibitem [{\citenamefont {Benzo}\ \emph {et~al.}(2019)\citenamefont {Benzo},
  \citenamefont {Combettes}, \citenamefont {Pecassou}, \citenamefont {Combe},
  \citenamefont {Benoit}, \citenamefont {Respaud},\ and\ \citenamefont
  {Casanove}}]{Benzo2019Sep}%
  \BibitemOpen
  \bibfield  {author} {\bibinfo {author} {\bibfnamefont {P.}~\bibnamefont
  {Benzo}}, \bibinfo {author} {\bibfnamefont {S.}~\bibnamefont {Combettes}},
  \bibinfo {author} {\bibfnamefont {B.}~\bibnamefont {Pecassou}}, \bibinfo
  {author} {\bibfnamefont {N.}~\bibnamefont {Combe}}, \bibinfo {author}
  {\bibfnamefont {M.}~\bibnamefont {Benoit}}, \bibinfo {author} {\bibfnamefont
  {M.}~\bibnamefont {Respaud}}, \ and\ \bibinfo {author} {\bibfnamefont
  {M.~J.}\ \bibnamefont {Casanove}},\ }\href {\doibase
  10.1103/PhysRevMaterials.3.096001} {\bibfield  {journal} {\bibinfo  {journal}
  {Phys. Rev. Mater.}\ }\textbf {\bibinfo {volume} {3}},\ \bibinfo {pages}
  {096001} (\bibinfo {year} {2019})}\BibitemShut {NoStop}%
\bibitem [{\citenamefont {Tymoczko}\ \emph {et~al.}(2019)\citenamefont
  {Tymoczko}, \citenamefont {Kamp}, \citenamefont {Rehbock}, \citenamefont
  {Kienle}, \citenamefont {Cattaruzza}, \citenamefont {Barcikowski},\ and\
  \citenamefont {Amendola}}]{Tymoczko2019}%
  \BibitemOpen
  \bibfield  {author} {\bibinfo {author} {\bibfnamefont {A.}~\bibnamefont
  {Tymoczko}}, \bibinfo {author} {\bibfnamefont {M.}~\bibnamefont {Kamp}},
  \bibinfo {author} {\bibfnamefont {C.}~\bibnamefont {Rehbock}}, \bibinfo
  {author} {\bibfnamefont {L.}~\bibnamefont {Kienle}}, \bibinfo {author}
  {\bibfnamefont {E.}~\bibnamefont {Cattaruzza}}, \bibinfo {author}
  {\bibfnamefont {S.}~\bibnamefont {Barcikowski}}, \ and\ \bibinfo {author}
  {\bibfnamefont {V.}~\bibnamefont {Amendola}},\ }\href {\doibase
  10.1039/C9NH00332K} {\bibfield  {journal} {\bibinfo  {journal} {Nanoscale
  Horiz.}\ }\textbf {\bibinfo {volume} {4}},\ \bibinfo {pages} {1326} (\bibinfo
  {year} {2019})}\BibitemShut {NoStop}%
\bibitem [{\citenamefont {Tymoczko}\ \emph {et~al.}(2018)\citenamefont
  {Tymoczko}, \citenamefont {Kamp}, \citenamefont {Prymak}, \citenamefont
  {Rehbock}, \citenamefont {Jakobi}, \citenamefont
  {Sch{\ifmmode\ddot{u}\else\"{u}\fi}rmann}, \citenamefont {Kienle},\ and\
  \citenamefont {Barcikowski}}]{Tymoczko2018Sep}%
  \BibitemOpen
  \bibfield  {author} {\bibinfo {author} {\bibfnamefont {A.}~\bibnamefont
  {Tymoczko}}, \bibinfo {author} {\bibfnamefont {M.}~\bibnamefont {Kamp}},
  \bibinfo {author} {\bibfnamefont {O.}~\bibnamefont {Prymak}}, \bibinfo
  {author} {\bibfnamefont {C.}~\bibnamefont {Rehbock}}, \bibinfo {author}
  {\bibfnamefont {J.}~\bibnamefont {Jakobi}}, \bibinfo {author} {\bibfnamefont
  {U.}~\bibnamefont {Sch{\ifmmode\ddot{u}\else\"{u}\fi}rmann}}, \bibinfo
  {author} {\bibfnamefont {L.}~\bibnamefont {Kienle}}, \ and\ \bibinfo {author}
  {\bibfnamefont {S.}~\bibnamefont {Barcikowski}},\ }\href {\doibase
  10.1039/C8NR03962C} {\bibfield  {journal} {\bibinfo  {journal} {Nanoscale}\
  }\textbf {\bibinfo {volume} {10}},\ \bibinfo {pages} {16434} (\bibinfo {year}
  {2018})}\BibitemShut {NoStop}%
\bibitem [{\citenamefont {Ponchet}\ \emph {et~al.}(2020)\citenamefont
  {Ponchet}, \citenamefont {Combettes}, \citenamefont {Benzo}, \citenamefont
  {Tarrat}, \citenamefont {Casanove},\ and\ \citenamefont
  {Benoit}}]{Ponchet2020Aug}%
  \BibitemOpen
  \bibfield  {author} {\bibinfo {author} {\bibfnamefont {A.}~\bibnamefont
  {Ponchet}}, \bibinfo {author} {\bibfnamefont {S.}~\bibnamefont {Combettes}},
  \bibinfo {author} {\bibfnamefont {P.}~\bibnamefont {Benzo}}, \bibinfo
  {author} {\bibfnamefont {N.}~\bibnamefont {Tarrat}}, \bibinfo {author}
  {\bibfnamefont {M.~J.}\ \bibnamefont {Casanove}}, \ and\ \bibinfo {author}
  {\bibfnamefont {M.}~\bibnamefont {Benoit}},\ }\href {\doibase
  10.1063/5.0014906} {\bibfield  {journal} {\bibinfo  {journal} {J. Appl.
  Phys.}\ }\textbf {\bibinfo {volume} {128}},\ \bibinfo {pages} {055307}
  (\bibinfo {year} {2020})}\BibitemShut {NoStop}%
\bibitem [{\citenamefont {Vernieres}\ \emph {et~al.}(2019)\citenamefont
  {Vernieres}, \citenamefont {Steinhauer}, \citenamefont {Zhao}, \citenamefont
  {Grammatikopoulos}, \citenamefont {Ferrando}, \citenamefont {Nordlund},
  \citenamefont {Djurabekova},\ and\ \citenamefont
  {Sowwan}}]{Vernieres2019Jul}%
  \BibitemOpen
  \bibfield  {author} {\bibinfo {author} {\bibfnamefont {J.}~\bibnamefont
  {Vernieres}}, \bibinfo {author} {\bibfnamefont {S.}~\bibnamefont
  {Steinhauer}}, \bibinfo {author} {\bibfnamefont {J.}~\bibnamefont {Zhao}},
  \bibinfo {author} {\bibfnamefont {P.}~\bibnamefont {Grammatikopoulos}},
  \bibinfo {author} {\bibfnamefont {R.}~\bibnamefont {Ferrando}}, \bibinfo
  {author} {\bibfnamefont {K.}~\bibnamefont {Nordlund}}, \bibinfo {author}
  {\bibfnamefont {F.}~\bibnamefont {Djurabekova}}, \ and\ \bibinfo {author}
  {\bibfnamefont {M.}~\bibnamefont {Sowwan}},\ }\href {\doibase
  10.1002/advs.201900447} {\bibfield  {journal} {\bibinfo  {journal} {Adv.
  Sci.}\ }\textbf {\bibinfo {volume} {6}},\ \bibinfo {pages} {1900447}
  (\bibinfo {year} {2019})}\BibitemShut {NoStop}%
\bibitem [{\citenamefont {Calvo}\ \emph {et~al.}(2017)\citenamefont {Calvo},
  \citenamefont {Combe}, \citenamefont {Morillo},\ and\ \citenamefont
  {Benoit}}]{Calvo2017Mar}%
  \BibitemOpen
  \bibfield  {author} {\bibinfo {author} {\bibfnamefont {F.}~\bibnamefont
  {Calvo}}, \bibinfo {author} {\bibfnamefont {N.}~\bibnamefont {Combe}},
  \bibinfo {author} {\bibfnamefont {J.}~\bibnamefont {Morillo}}, \ and\
  \bibinfo {author} {\bibfnamefont {M.}~\bibnamefont {Benoit}},\ }\href
  {\doibase 10.1021/acs.jpcc.6b12551} {\bibfield  {journal} {\bibinfo
  {journal} {J. Phys. Chem. C}\ }\textbf {\bibinfo {volume} {121}},\ \bibinfo
  {pages} {4680} (\bibinfo {year} {2017})}\BibitemShut {NoStop}%
\bibitem [{\citenamefont {Hong}\ and\ \citenamefont
  {Rahman}(2015)}]{Hong2015Oct}%
  \BibitemOpen
  \bibfield  {author} {\bibinfo {author} {\bibfnamefont {S.}~\bibnamefont
  {Hong}}\ and\ \bibinfo {author} {\bibfnamefont {T.~S.}\ \bibnamefont
  {Rahman}},\ }\href {\doibase 10.1039/C5CP00299K} {\bibfield  {journal}
  {\bibinfo  {journal} {Phys. Chem. Chem. Phys.}\ }\textbf {\bibinfo {volume}
  {17}},\ \bibinfo {pages} {28177} (\bibinfo {year} {2015})}\BibitemShut
  {NoStop}%
\bibitem [{\citenamefont {Combettes}\ \emph {et~al.}(2020)\citenamefont
  {Combettes}, \citenamefont {Lam}, \citenamefont {Benzo}, \citenamefont
  {Ponchet}, \citenamefont {Casanove}, \citenamefont {Calvo},\ and\
  \citenamefont {Benoit}}]{Combettes2020Sep}%
  \BibitemOpen
  \bibfield  {author} {\bibinfo {author} {\bibfnamefont {S.}~\bibnamefont
  {Combettes}}, \bibinfo {author} {\bibfnamefont {J.}~\bibnamefont {Lam}},
  \bibinfo {author} {\bibfnamefont {P.}~\bibnamefont {Benzo}}, \bibinfo
  {author} {\bibfnamefont {A.}~\bibnamefont {Ponchet}}, \bibinfo {author}
  {\bibfnamefont {M.-J.}\ \bibnamefont {Casanove}}, \bibinfo {author}
  {\bibfnamefont {F.}~\bibnamefont {Calvo}}, \ and\ \bibinfo {author}
  {\bibfnamefont {M.}~\bibnamefont {Benoit}},\ }\href {\doibase
  10.1039/D0NR04425C} {\bibfield  {journal} {\bibinfo  {journal} {Nanoscale}\
  }\textbf {\bibinfo {volume} {12}},\ \bibinfo {pages} {18079} (\bibinfo {year}
  {2020})}\BibitemShut {NoStop}%
\bibitem [{\citenamefont {Zhou}\ \emph {et~al.}(2004)\citenamefont {Zhou},
  \citenamefont {Johnson},\ and\ \citenamefont {Wadley}}]{Zhou2004Apr}%
  \BibitemOpen
  \bibfield  {author} {\bibinfo {author} {\bibfnamefont {X.~W.}\ \bibnamefont
  {Zhou}}, \bibinfo {author} {\bibfnamefont {R.~A.}\ \bibnamefont {Johnson}}, \
  and\ \bibinfo {author} {\bibfnamefont {H.~N.~G.}\ \bibnamefont {Wadley}},\
  }\href {\doibase 10.1103/PhysRevB.69.144113} {\bibfield  {journal} {\bibinfo
  {journal} {Phys. Rev. B}\ }\textbf {\bibinfo {volume} {69}},\ \bibinfo
  {pages} {144113} (\bibinfo {year} {2004})}\BibitemShut {NoStop}%
\bibitem [{\citenamefont {Takahashi}\ \emph {et~al.}(2017)\citenamefont
  {Takahashi}, \citenamefont {Seko},\ and\ \citenamefont
  {Tanaka}}]{Takahashi2017Nov}%
  \BibitemOpen
  \bibfield  {author} {\bibinfo {author} {\bibfnamefont {A.}~\bibnamefont
  {Takahashi}}, \bibinfo {author} {\bibfnamefont {A.}~\bibnamefont {Seko}}, \
  and\ \bibinfo {author} {\bibfnamefont {I.}~\bibnamefont {Tanaka}},\ }\href
  {\doibase 10.1103/PhysRevMaterials.1.063801} {\bibfield  {journal} {\bibinfo
  {journal} {Phys. Rev. Materials}\ }\textbf {\bibinfo {volume} {1}},\ \bibinfo
  {pages} {063801} (\bibinfo {year} {2017})}\BibitemShut {NoStop}%
\bibitem [{\citenamefont {Seko}\ \emph {et~al.}(2019)\citenamefont {Seko},
  \citenamefont {Togo},\ and\ \citenamefont {Tanaka}}]{Seko2019Jun}%
  \BibitemOpen
  \bibfield  {author} {\bibinfo {author} {\bibfnamefont {A.}~\bibnamefont
  {Seko}}, \bibinfo {author} {\bibfnamefont {A.}~\bibnamefont {Togo}}, \ and\
  \bibinfo {author} {\bibfnamefont {I.}~\bibnamefont {Tanaka}},\ }\href
  {\doibase 10.1103/PhysRevB.99.214108} {\bibfield  {journal} {\bibinfo
  {journal} {Phys. Rev. B}\ }\textbf {\bibinfo {volume} {99}},\ \bibinfo
  {pages} {214108} (\bibinfo {year} {2019})}\BibitemShut {NoStop}%
\bibitem [{\citenamefont {Wood}\ and\ \citenamefont
  {Thompson}(2018)}]{Wood2018Jun}%
  \BibitemOpen
  \bibfield  {author} {\bibinfo {author} {\bibfnamefont {M.~A.}\ \bibnamefont
  {Wood}}\ and\ \bibinfo {author} {\bibfnamefont {A.~P.}\ \bibnamefont
  {Thompson}},\ }\href {\doibase 10.1063/1.5017641@jcp.2018.DETC2018.issue-1}
  {\bibfield  {journal} {\bibinfo  {journal} {J. Chem. Phys.}\ }\textbf
  {\bibinfo {volume} {DETC2018}},\ \bibinfo {pages} {241721} (\bibinfo {year}
  {2018})}\BibitemShut {NoStop}%
\bibitem [{\citenamefont {Wang}\ \emph {et~al.}(2019)\citenamefont {Wang},
  \citenamefont {Guo}, \citenamefont {Zhang}, \citenamefont {Wang},\ and\
  \citenamefont {Xue}}]{Wang2019Jun}%
  \BibitemOpen
  \bibfield  {author} {\bibinfo {author} {\bibfnamefont {H.}~\bibnamefont
  {Wang}}, \bibinfo {author} {\bibfnamefont {X.}~\bibnamefont {Guo}}, \bibinfo
  {author} {\bibfnamefont {L.}~\bibnamefont {Zhang}}, \bibinfo {author}
  {\bibfnamefont {H.}~\bibnamefont {Wang}}, \ and\ \bibinfo {author}
  {\bibfnamefont {J.}~\bibnamefont {Xue}},\ }\href {\doibase 10.1063/1.5098061}
  {\bibfield  {journal} {\bibinfo  {journal} {Appl. Phys. Lett.}\ }\textbf
  {\bibinfo {volume} {114}},\ \bibinfo {pages} {244101} (\bibinfo {year}
  {2019})}\BibitemShut {NoStop}%
\bibitem [{\citenamefont {Stillinger}\ and\ \citenamefont
  {Weber}(1985)}]{Stillinger1985Apr}%
  \BibitemOpen
  \bibfield  {author} {\bibinfo {author} {\bibfnamefont {F.~H.}\ \bibnamefont
  {Stillinger}}\ and\ \bibinfo {author} {\bibfnamefont {T.~A.}\ \bibnamefont
  {Weber}},\ }\href {\doibase 10.1103/PhysRevB.31.5262} {\bibfield  {journal}
  {\bibinfo  {journal} {Phys. Rev. B}\ }\textbf {\bibinfo {volume} {31}},\
  \bibinfo {pages} {5262} (\bibinfo {year} {1985})}\BibitemShut {NoStop}%
\bibitem [{\citenamefont {Daw}\ and\ \citenamefont
  {Baskes}(1984)}]{Daw1984Jun}%
  \BibitemOpen
  \bibfield  {author} {\bibinfo {author} {\bibfnamefont {M.~S.}\ \bibnamefont
  {Daw}}\ and\ \bibinfo {author} {\bibfnamefont {M.~I.}\ \bibnamefont
  {Baskes}},\ }\href {\doibase 10.1103/PhysRevB.29.6443} {\bibfield  {journal}
  {\bibinfo  {journal} {Phys. Rev. B}\ }\textbf {\bibinfo {volume} {29}},\
  \bibinfo {pages} {6443} (\bibinfo {year} {1984})}\BibitemShut {NoStop}%
\bibitem [{\citenamefont {Efron}\ \emph {et~al.}(2004)\citenamefont {Efron},
  \citenamefont {Hastie}, \citenamefont {Johnstone},\ and\ \citenamefont
  {Tibshirani}}]{Efron2004Apr}%
  \BibitemOpen
  \bibfield  {author} {\bibinfo {author} {\bibfnamefont {B.}~\bibnamefont
  {Efron}}, \bibinfo {author} {\bibfnamefont {T.}~\bibnamefont {Hastie}},
  \bibinfo {author} {\bibfnamefont {I.}~\bibnamefont {Johnstone}}, \ and\
  \bibinfo {author} {\bibfnamefont {R.}~\bibnamefont {Tibshirani}},\ }\href
  {\doibase 10.1214/009053604000000067} {\bibfield  {journal} {\bibinfo
  {journal} {Annals of Statistics}\ }\textbf {\bibinfo {volume} {32}},\
  \bibinfo {pages} {407} (\bibinfo {year} {2004})}\BibitemShut {NoStop}%
\bibitem [{\citenamefont {Hung}\ \emph {et~al.}(2015)\citenamefont {Hung},
  \citenamefont {Hue},\ and\ \citenamefont {Duc}}]{MaP}%
  \BibitemOpen
  \bibfield  {author} {\bibinfo {author} {\bibfnamefont {N.}~\bibnamefont
  {Hung}}, \bibinfo {author} {\bibfnamefont {T.}~\bibnamefont {Hue}}, \ and\
  \bibinfo {author} {\bibfnamefont {N.}~\bibnamefont {Duc}},\ }\href
  {https://js.vnu.edu.vn/MaP/article/view/114} {\bibfield  {journal} {\bibinfo
  {journal} {VNU Journal of Science: Mathematics - Physics}\ }\textbf {\bibinfo
  {volume} {31}} (\bibinfo {year} {2015})}\BibitemShut {NoStop}%
\bibitem [{\citenamefont {Kresse}\ and\ \citenamefont
  {Furthm\"{u}ller}(1996)}]{kresse_efficiency_1996}%
  \BibitemOpen
  \bibfield  {author} {\bibinfo {author} {\bibfnamefont {G.}~\bibnamefont
  {Kresse}}\ and\ \bibinfo {author} {\bibfnamefont {J.}~\bibnamefont
  {Furthm\"{u}ller}},\ }\href {\doibase DOI: 10.1016/0927-0256(96)00008-0}
  {\bibfield  {journal} {\bibinfo  {journal} {Computational Materials Science}\
  }\textbf {\bibinfo {volume} {6}},\ \bibinfo {pages} {15 } (\bibinfo {year}
  {1996})}\BibitemShut {NoStop}%
\bibitem [{\citenamefont {Bl\"{o}chl}(1994)}]{blochl_projector_1994}%
  \BibitemOpen
  \bibfield  {author} {\bibinfo {author} {\bibfnamefont {P.~E.}\ \bibnamefont
  {Bl\"{o}chl}},\ }\href {\doibase 10.1103/PhysRevB.50.17953} {\bibfield
  {journal} {\bibinfo  {journal} {Phys. Rev. B}\ }\textbf {\bibinfo {volume}
  {50}},\ \bibinfo {pages} {17953} (\bibinfo {year} {1994})}\BibitemShut
  {NoStop}%
\bibitem [{\citenamefont {Zuo}\ \emph {et~al.}(2020)\citenamefont {Zuo},
  \citenamefont {Chen}, \citenamefont {Li}, \citenamefont {Deng}, \citenamefont
  {Chen}, \citenamefont {Behler}, \citenamefont
  {Cs{\ifmmode\acute{a}\else\'{a}\fi}nyi}, \citenamefont {Shapeev},
  \citenamefont {Thompson}, \citenamefont {Wood},\ and\ \citenamefont
  {Ong}}]{Zuo2020Jan}%
  \BibitemOpen
  \bibfield  {author} {\bibinfo {author} {\bibfnamefont {Y.}~\bibnamefont
  {Zuo}}, \bibinfo {author} {\bibfnamefont {C.}~\bibnamefont {Chen}}, \bibinfo
  {author} {\bibfnamefont {X.}~\bibnamefont {Li}}, \bibinfo {author}
  {\bibfnamefont {Z.}~\bibnamefont {Deng}}, \bibinfo {author} {\bibfnamefont
  {Y.}~\bibnamefont {Chen}}, \bibinfo {author} {\bibfnamefont {J.}~\bibnamefont
  {Behler}}, \bibinfo {author} {\bibfnamefont {G.}~\bibnamefont
  {Cs{\ifmmode\acute{a}\else\'{a}\fi}nyi}}, \bibinfo {author} {\bibfnamefont
  {A.~V.}\ \bibnamefont {Shapeev}}, \bibinfo {author} {\bibfnamefont {A.~P.}\
  \bibnamefont {Thompson}}, \bibinfo {author} {\bibfnamefont {M.~A.}\
  \bibnamefont {Wood}}, \ and\ \bibinfo {author} {\bibfnamefont {S.~P.}\
  \bibnamefont {Ong}},\ }\href {\doibase 10.1021/acs.jpca.9b08723} {\bibfield
  {journal} {\bibinfo  {journal} {J. Phys. Chem. A}\ }\textbf {\bibinfo
  {volume} {124}},\ \bibinfo {pages} {731} (\bibinfo {year}
  {2020})}\BibitemShut {NoStop}%
\bibitem [{\citenamefont {Benoit}\ \emph {et~al.}(2012)\citenamefont {Benoit},
  \citenamefont {Langlois}, \citenamefont {Combe}, \citenamefont {Tang},\ and\
  \citenamefont {Casanove}}]{Benoit2012Aug}%
  \BibitemOpen
  \bibfield  {author} {\bibinfo {author} {\bibfnamefont {M.}~\bibnamefont
  {Benoit}}, \bibinfo {author} {\bibfnamefont {C.}~\bibnamefont {Langlois}},
  \bibinfo {author} {\bibfnamefont {N.}~\bibnamefont {Combe}}, \bibinfo
  {author} {\bibfnamefont {H.}~\bibnamefont {Tang}}, \ and\ \bibinfo {author}
  {\bibfnamefont {M.-J.}\ \bibnamefont {Casanove}},\ }\href {\doibase
  10.1103/PhysRevB.86.075460} {\bibfield  {journal} {\bibinfo  {journal} {Phys.
  Rev. B}\ }\textbf {\bibinfo {volume} {86}},\ \bibinfo {pages} {075460}
  (\bibinfo {year} {2012})}\BibitemShut {NoStop}%
\bibitem [{\citenamefont {Plimpton}(1995)}]{plimpton1995}%
  \BibitemOpen
  \bibfield  {author} {\bibinfo {author} {\bibfnamefont {S.}~\bibnamefont
  {Plimpton}},\ }\href {http://lammps.sandia.gov/index.html} {\bibfield
  {journal} {\bibinfo  {journal} {J. Comput. Phys.}\ }\textbf {\bibinfo
  {volume} {117}},\ \bibinfo {pages} {1} (\bibinfo {year} {1995})}\BibitemShut
  {NoStop}%
\bibitem [{\citenamefont {Togo}\ and\ \citenamefont {Tanaka}(2015)}]{phonopy}%
  \BibitemOpen
  \bibfield  {author} {\bibinfo {author} {\bibfnamefont {A.}~\bibnamefont
  {Togo}}\ and\ \bibinfo {author} {\bibfnamefont {I.}~\bibnamefont {Tanaka}},\
  }\href@noop {} {\bibfield  {journal} {\bibinfo  {journal} {Scr. Mater.}\
  }\textbf {\bibinfo {volume} {108}},\ \bibinfo {pages} {1} (\bibinfo {year}
  {2015})}\BibitemShut {NoStop}%
\bibitem [{\citenamefont {Byggm{\ifmmode\ddot{a}\else\"{a}\fi}star}\ \emph
  {et~al.}(2019)\citenamefont {Byggm{\ifmmode\ddot{a}\else\"{a}\fi}star},
  \citenamefont {Hamedani}, \citenamefont {Nordlund},\ and\ \citenamefont
  {Djurabekova}}]{Byggmastar2019Oct}%
  \BibitemOpen
  \bibfield  {author} {\bibinfo {author} {\bibfnamefont {J.}~\bibnamefont
  {Byggm{\ifmmode\ddot{a}\else\"{a}\fi}star}}, \bibinfo {author} {\bibfnamefont
  {A.}~\bibnamefont {Hamedani}}, \bibinfo {author} {\bibfnamefont
  {K.}~\bibnamefont {Nordlund}}, \ and\ \bibinfo {author} {\bibfnamefont
  {F.}~\bibnamefont {Djurabekova}},\ }\href {\doibase
  10.1103/PhysRevB.100.144105} {\bibfield  {journal} {\bibinfo  {journal}
  {Phys. Rev. B}\ }\textbf {\bibinfo {volume} {100}},\ \bibinfo {pages}
  {144105} (\bibinfo {year} {2019})}\BibitemShut {NoStop}%
\bibitem [{\citenamefont {Rushton}()}]{atsim.potentials}%
  \BibitemOpen
  \bibfield  {author} {\bibinfo {author} {\bibfnamefont {M.}~\bibnamefont
  {Rushton}},\ }\href {https://atsimpotentials.readthedocs.io/} {\enquote
  {\bibinfo {title} {atsim.potentials - potential model tabulation for atomic
  scale simulation},}\ }\BibitemShut {NoStop}%
\end{thebibliography}
\end{document}